\documentclass[final,5p,times,twocolumn]{elsarticle}
\usepackage{ifpdf}
\usepackage[usenames]{color}
\usepackage{graphicx}
\usepackage{amssymb}
\usepackage{amsmath}
\usepackage{ulem}
\journal{Physics Letters A}
\begin{document}
\begin{frontmatter}
\title{Modulational Instabity of Spin-Orbit Coupled Bose-Einstein Condensates \\ in Discrete Media}
\author{ S. Sabari$^{1,*}$, R. TamilThiruvalluvar$^{2,\dag}$ and R. Radha$^{1,\#}$}
\address{$^{1}$Centre for Nonlinear Science, Department of Physics, Government College for Women(A), Kumbakonam 612001, India\\$^{2}$Department of Physics, Pondicherry University, Puducherry - 605014, India\\$^*$ssabari01@gmail.com,\,\,$^\dag$taphy1905@gmail.com,\,\, $^\#$vittal.cnls@gmail.com}
%
\date{\today}
\begin{abstract}
We address the impact of intra-site spin-orbit  (SO) coupling and associated inter-component Rabi coupling on the modulational instability (MI) of plane-wave states in two-component discrete Bose-Einstein condensates (BECs). Conditions for the onset of the MI and the respective instability are found analytically. SO coupling allows us to produce the MI even for a small initial wavenumber ($q<\pi /2$) for  miscible states. In particular, SO coupling introduces MI even in the absence of hopping coefficient, a concept  which may have wider ramifications in heavy atomic BECs. 
We have also shown how our results of the linear stability analysis can be corroborated numerically. The fact that we have brought out the stability criteria in different domains of  system parameters means that our model is tailor made experiments.
\end{abstract}


\end{frontmatter}

\date{\today}
\section{Introduction}
\label{sec1} 
Even though the ubiquitous nature of the spin-orbit (SO) coupling, which involves the interaction between quantum particle's spin and its momentum is quite well known, it came to the fore only after the experimental realization of Bose-Einstein condensates (BECs), a new state of mater comprising  of ultracold atoms getting piled up in the ground state like a giant matter wave. It must be emphasized  that BECs provide  an ideal platform (launching pad) to engineer SO coupling in a neutral atomic BEC by dressing two atomic spin states with parts of laser. In fact, in the presence of laser coupling~\cite{Spielman}, the interaction between the two dressed atomic spin states are modified driving a quantum phase transition from a spatially spin mixed state (laser off) to a phase separated state(stripped phase). The stripped phase which occurs above a critical laser density initiates a phase transition by engineering SO coupling thereby leading  to the rich possibility of generating synthetic electric and magnetic fields. Recent investigations have explored the possibility of identifying tunable spin-orbit coupled (SOC) BECs with various trapping potentials and stable domains of condensates have been observed. 

The SOC-BECs involving the interplay between the nonlinear two-body interaction with the linear interaction of SOC have been investigated in the last few years. For instance, it has been  reported that the new ground-state phase can be created in two-component nonlinear systems such as stripes, phase separation,etc~\cite{Wang}, tricritical points~\cite{Li}, different type of solitons~\cite{Achilleos,Kartashov,Lobanov,Xu}, 2D solitons with embedded vorticity and vortex lattice~\cite{Sakaguchi-Li-Malomed1,Salasnich-Malomed1,Sakaguchi-Li-Malomed2,Sakaguchi-Meqs}.

The combinations of optical lattice (OL) potential with the interaction of SO coupling was shown to posses interesting phenomena like flattening of Bloch potential~\cite{YZhang}, atomic Zitterbewegung~\cite{Larson} and new topological phases~\cite{Stanescu}. Generally, BECs become fragmented on each lattice when it is trapped by deep OL potential. Such a system has been effectively described in the tight-binding approximation by discrete version of the corresponding Gross-Pitaevskii (GP) equation~\cite{Trombettoni,Kevrekidis,Gligoric}. Recent investigations on composite solitons and localized modes are carried out on Rashba type SOC-BEC trapped in deep OL~\cite{Sakaguchi-Meqs,Belicev}.

The modulational instability (MI) is a generic phenomenon which occurs in a dynamical system like fluid dynamics, nonlinear optics, plasma physics due to the interplay between nonlinearity and dispersion (or diffraction in spatial domain) resulting in the fragmentation of carrier waves into a train of localized waves~\cite{Benjamin}. 
Recently, the onset of MI in scalar~\cite{Al Khawaja,K.E.Strecker,Sabari} and vector BECs~\cite{Goldstein,Kasamatsu1,Kasamatsu2} within the framework of continuous NLS type (GP) equations has been investigated. The above scenario could drastically change in discrete multicomponent settings~\cite{Kobyakov 2C MI,ZRaptiMI,Baizakov 2C MI,Guang 2C MI,Ruostekoski}. 

It has been found that the onset of MI not only depends on the sign of diffraction (or dispersion), but also on the sign of inter-species and intra-species interaction strengths. Recently, MI in SOC-BECs has been investigated in continuous media~\cite{ishfaq,mithun}. However, the condition for the onset MI in discrete media (or settings) has not yet been {\it completely} brought out, particularly after the advent of SOC-BECs. 
In the light of the above developments, it would be interesting to study the onset of MI in SOC-BEC in discrete media. The freedom associated with the SOC-BECs by virtue of Rasbha coupling, SO coupling, intra/inter species interaction and dispersion could lead to the identification of several interesting stable domains from an experimental perspective. 

 In view of the above, we have made an attempt to analyze the impact of SO coupling on MI in discrete BECs. This paper is structured as follows. In sec. 2, the SO coupling model is constructed in terms of coupled discrete GP equations and associate dispersion relations are found via linear stability analysis. The results of semi analytical  analysis are discussed with different signs of intra- and inter component fragmentation with SO coupling interaction in sec. 3. Finally in sec. 4, we elaborate on the salient features of the present investigation.

\section{Theoretical model and dispersion relation for MI}

\label{sec2} We consider a BEC comprising of atoms in different hyperfine ground states corresponding to $|F=1,m_{F}=0\rangle $ and $|F=1,m_{F}=-1\rangle $~\cite{Spielman}. These settings induce the SO coupling resulting in the formation of two dressed states as a basis. The dynamics of such a system described by mean-field GP equation in the presence of deep OL in spinor form is governed  by
\begin{equation}
\begin{split}
\imath \hbar \frac{\partial \Psi }{\partial t}=\left[ \frac{\hat{p}^2}{2m}%
+V_{OL}+\hat{H}_{SOC}+\hat{H}_{mix}+H_{int}\right] \Psi
\end{split}
\end{equation}%
where $\Psi =(\Psi ^{+},\Psi ^{-})^{T}$ represents the spinor wave function in the dressed atomic state and the term $\frac{\hat{p}^2}{2m}$ represents the kinetic energy of the quasi-one-dimensional condensate along the x-direction. The Hamiltonian term which describes the SOC interaction has the following form,
\begin{equation}
\hat{H}_{SOC}=-\imath \frac{\hbar ^{2}}{m}\gamma \sigma _{z}\frac{\partial }{%
\partial x}
\end{equation}%
where $\gamma$ is the strength of SO coupling and $\sigma _{z}$ is z-component Pauli's spin matrix. The linear coupling term is represented by $\hat{H}_{mix}=\hbar \delta \sigma _{x}$, which orginates through magnetic field in intra-SOC system where $\delta $ is strength of Rabi coupling.  $H_{int}$ accounts for the intra- and inter-component interactions in the BEC for the mixture which are nonlinear in nature.

Such a system in deep OL leads to the replacement of the continuous GP equation by the discrete counterpart to single component BECs in earlier studies~\cite{Trombettoni,Kevrekidis,Gligoric} and two-component BECs~\cite{ZRaptiMI,Baizakov 2C MI,Guang 2C MI,Ruostekoski}. Thus, we arrive at a system of coupled discrete GP equation with intra-SOC and Rabi coupling of the following form ~\cite{Sakaguchi-Meqs,Belicev}%
,
\begin{equation}
\begin{split}
i \frac{\partial \Psi_{n}^{\pm}}{\partial t}=& - \Gamma^{\pm}\big(%
\Psi_{n+1}^{\pm} + \Psi_{n-1}^{\pm} - 2\Psi_{n}^{\pm} \big) \mp \imath \gamma \big(\Psi_{n+1}^{\pm} - \Psi_{n-1}^{\pm}\big)\\
& + \big(g^{\pm}|\Psi_{n}^{\pm}|^{2} + G^{c\pm}|\Psi_{n}^{\mp}|^{2}\big)\Psi_{n}^{\pm}-\delta
\Psi_{n}^{\mp}. 
\end{split}
\label{dgpe}
\end{equation}%
where, $\Gamma $ represents hopping coefficient of adjacent lattice sites while  $g^{\pm }$ and $G^{c\pm }$ represent intra- and inter-component collision strengths
respectively. This system conserves two quantities for infinite lattices namely
\begin{equation*}
N=\sum_{n}\left( |\Psi _{n}^{+}|^{2}+|\Psi _{n}^{-}|^{2}\right)
\end{equation*}%
and the Hamiltonian which is
\begin{equation}
\begin{split}
H&= \sum_{n}\Big(-\Gamma \left(\Psi _{n}^{+\ast }\Psi _{n+1}^{+}+\Psi _{n}^{-\ast
}\Psi _{n+1}^{-}\right)-\imath \gamma (\Psi _{n}^{+\ast }\Psi _{n+1}^{+} -\\
& \Psi _{n}^{-\ast }\Psi _{n+1}^{-})+\frac{1}{4}\left(g^{+}|\Psi
_{n}^{+}|^{4}+g^{-}|\Psi _{n}^{-}|^{4}\right)+\frac{1}{4}(G^{c+}|\Psi
_{n}^{+}|^{2} \\
& |\Psi _{n}^{-}|^{2}+G^{c-}|\Psi _{n}^{+}|^{2}|\Psi _{n}^{-}|^{2})-\delta
\Psi _{n}^{+\ast }\Psi _{n}^{-}\Big)+c.c.
\end{split}%
\label{hamil}
\end{equation}%

We now explore the stability of stationary plane-wave solutions of Equ.(\ref{dgpe}) with the following ansatz
\begin{equation}
\Psi _{n}^{\pm }=u^{\pm }\exp [\imath \omega ^{\pm }t+\imath nq^{\pm }].
\label{ansat}
\end{equation}
Employing the ansatz given by Equ.(\ref{ansat}), one can obtain the following dispersion relation
\begin{equation}
\omega =-2\Gamma _{\pm }\cos (q)+g^{\pm }u^{\pm 2}+G^{c\pm }u_{\mp
}^{2}-\frac{u^{\mp }}{u^{\pm }}\delta \pm 2\gamma \sin (q)
\label{disp}
\end{equation}%
Next, we slightly modify the plane wave ansatz given by Equ.(\ref{ansat}) of the  following form 
\begin{equation*}
\Psi _{n}^{\pm }=(u^{\pm }+\xi ^{\pm })\exp [\imath \omega ^{\pm }t+\imath
nq^{\pm }]
\end{equation*}%
Subsequently, perturbed solution of the above ansatz was substituted in Equ.(\ref{dgpe}) to extract the discrete differential equation only in linear terms of $\xi ^{\pm }$ ($\xi ^{\pm \ast }$ is complex conjugate) as
\begin{equation}
\begin{split}
\imath \frac{\partial \xi _{n}^{\pm
 }}{\partial t}&= \,-\Gamma ^{\pm }\big(%
\xi _{n+1}^{\pm }+\xi _{n-1}^{\pm }-2\xi _{n}^{\pm }\big)+g^{\pm }(\xi
_{n}^{\pm }+\xi _{n}^{\pm \ast }) \\
& +G^{c\pm }(\xi _{n}^{\mp }+\xi _{n}^{\mp \ast })\mp \imath \gamma (\xi
_{n+1}^{\pm }-\xi _{n-1}^{\pm })-\delta \xi _{n}^{\mp }. \label{equ7}
\end{split}%
\end{equation}%
\color{black}
The perturbed amplitudes are taken as plane waves in discrete coordinates of the following form
\begin{equation}
\begin{split}
\xi _{n}^{+}& =\psi 1\exp [\imath \Omega t+\imath nQ]+\psi 2\exp [-\imath
\Omega t-\imath nQ] \\
\xi _{n}^{-}& =\phi 1\exp [\imath \Omega t+\imath nQ]+\phi 2\exp [-\imath
\Omega t-\imath nQ] \label{equ8}
\end{split}%
\end{equation}%
A straightforward substitution of equation (\ref{equ8}) in (\ref{equ7}) yields the following dispersion relation for eigenfrequency $\Omega$ as

\begin{align}
\begin{split}
\Omega^4+P_3\Omega^3+P_2\Omega^2+P_1\Omega+P_0=0  \label{Jaco9}
\end{split}%
\end{align}

\noindent where
\begin{align}
\begin{split}
P_0=&\big[(g+\gamma)(g+2G-\gamma)+\Lambda_+(\gamma-G)-\Lambda_-(G-\gamma+\Lambda_+)\big] \\&\big[(g-\gamma)(g-2G+\gamma)+\Lambda_-(G-\gamma)-\Lambda_+(\gamma-G+\Lambda_-)\big] \\
P_1=&2(\Lambda_--\Lambda_+)(\Lambda_-\Lambda_+-g^2+\gamma(\gamma-2G)) \\
P_2=&2g^2+4G\gamma-2\gamma^2+\Lambda_-^2+\Lambda_+^2-4\Lambda_-\Lambda_+ \\
P_3=&2(\Lambda_+-\Lambda_-)  \label{Jaco10}
\end{split}%
\end{align}
and
$\Lambda_\pm=\theta_\pm+g+\gamma$ and $\theta_\pm =2\cos(q)(\Gamma-\Gamma\cos(Q)\pm\gamma\sin(Q))$
%
The instability growth rate is defined as
$\zeta=2|\text{Im} (\Omega_\pm)|$

\section{Results And Discussion}
\label{sec3}

\subsection{MI in unstaggered mode}
In this section, we study the onset of MI in a unstaggered mode. It has been found that the MI gives rise to  unstaggered mode at $g^+>0$ ($q=0$), and staggered mode for $g^+<0$ ($q=\pi$). We have brought out the onset of MI domains in the absence/presence/ of all system parameters in Table 1.

\begin{table}[h!]
\caption {Review of the MI in discrete SOC-BECs at unstaggered mode}
\label{table2}
\begin{center}
\begin{tabular}{|l|l|l|l|l|}
\hline 
\textbf{Hopping }& &\textbf{Rabi } & \textbf{Possible} &  \\
\textbf{Strength}&\textbf{SOC} &\textbf{coup.}& \textbf{Cases} & \textbf{Conjecture} \\
\hline \hline
& & &$g>0$, $g_{12}>0$  & \,\,\,\,Unstable \\ 
& &  \,\,$\delta=0$ & $g>0$, $g_{12}<0$ & \,\,\,\,Unstable \\ 
& & & $g<0$, $g_{12}>0$ & \,\,\,\,Stable\\ 
&\,$\gamma=0$  & & $g<0$, $g_{12}<0$ & \,\,\,\,Stable \\  \cline{3-5}
& & &$g>0$, $g_{12}>0$  & \,\,\,\,Unstable \\ 
& &  \,\,$\delta=1$ & $g>0$, $g_{12}<0$ & \,\,\,\,Unstable \\ 
& & & $g<0$, $g_{12}>0$ & \,\,\,\,Stable\\ 
\,\,\,\,\,$\Gamma=1$ & & & $g<0$, $g_{12}<0$ & \,\,\,\,Unstable \\  
\cline{2-5}
& & &$g>0$, $g_{12}>0$  & \,\,\,\,Unstable \\ 
& &  \,\,$\delta=0$ & $g>0$, $g_{12}<0$ & \,\,\,\,Unstable \\ 
& & & $g<0$, $g_{12}>0$ & \,\,\,\,Stable\\ 
&\,$\gamma=1$  & & $g<0$, $g_{12}<0$ & \,\,\,\,Stable \\  \cline{3-5}
& & &$g>0$, $g_{12}>0$  & \,\,  Unstable\\ 
& &  \,\,$\delta=1$ & $g>0$, $g_{12}<0$ & \,\,\,\,Unstable \\ 
& & & $g<0$, $g_{12}>0$ & \,\, Stable\\ 
& & & $g<0$, $g_{12}<0$ & \,\,  Unstable\\  
\cline{1-5}
& & &$g>0$, $g_{12}>0$  & \,\, \\ 
& &  \,\,$\delta=0$ & $g>0$, $g_{12}<0$ & \,\,\,\,Unstable \\ 
& & & $g<0$, $g_{12}>0$ & \,\, \\ 
\,\,\,\,\,$\Gamma=0$ &\,$\gamma=1$  & & $g<0$, $g_{12}<0$ & \,\,  \\  \cline{3-5}
& & &$g>0$, $g_{12}>0$  & \,\,\,\,Unstable \\ 
& &  \,\,$\delta=1$ & $g>0$, $g_{12}<0$ & \,\,\,\,Unstable\\ 
& & & $g<0$, $g_{12}>0$ & \,\,\,\,Stable\\ 
& & & $g<0$, $g_{12}<0$ & \,\,\,\,Stable \\  
\hline
\end{tabular}
\end{center}
\end{table}

\begin{figure}[htbp!]
\includegraphics[width=.235\textwidth]{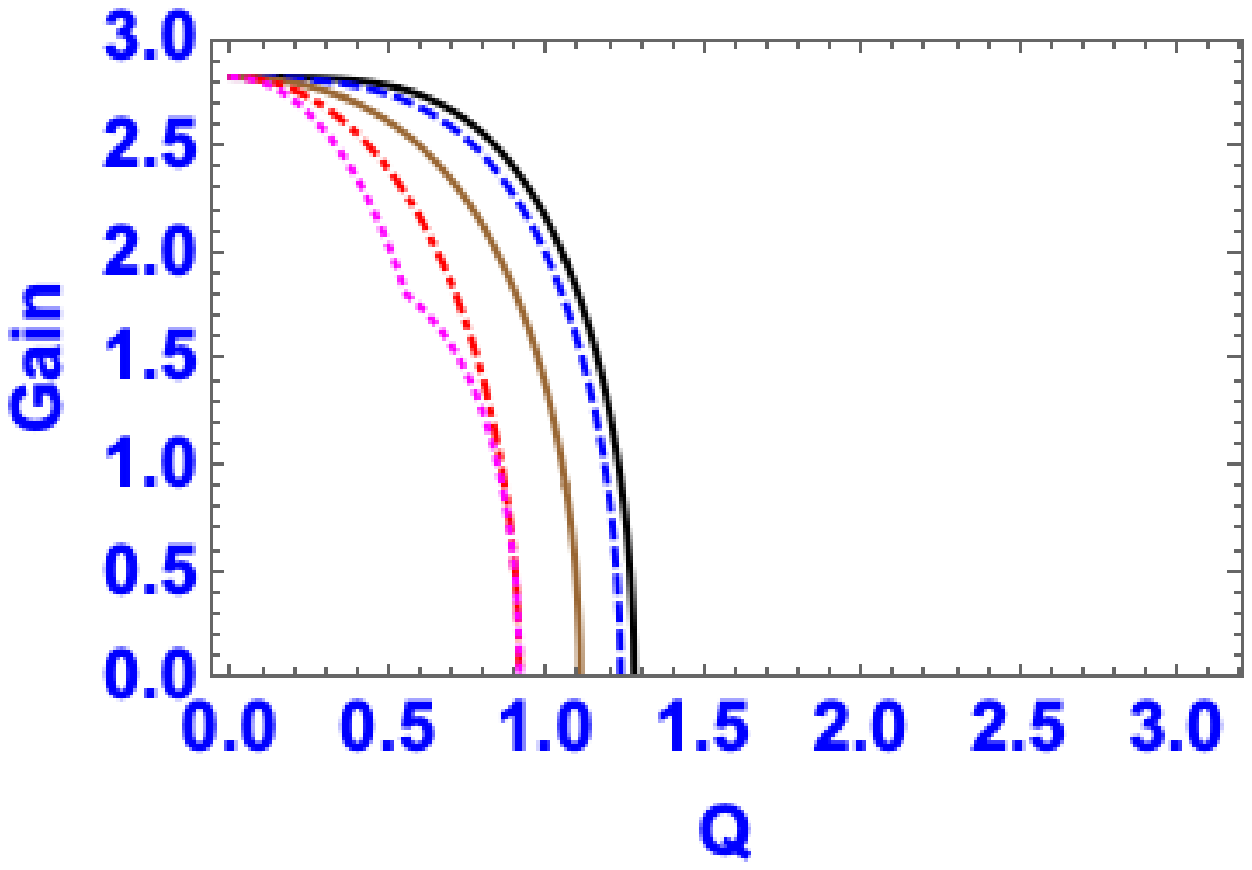} %
\includegraphics[width=.235\textwidth]{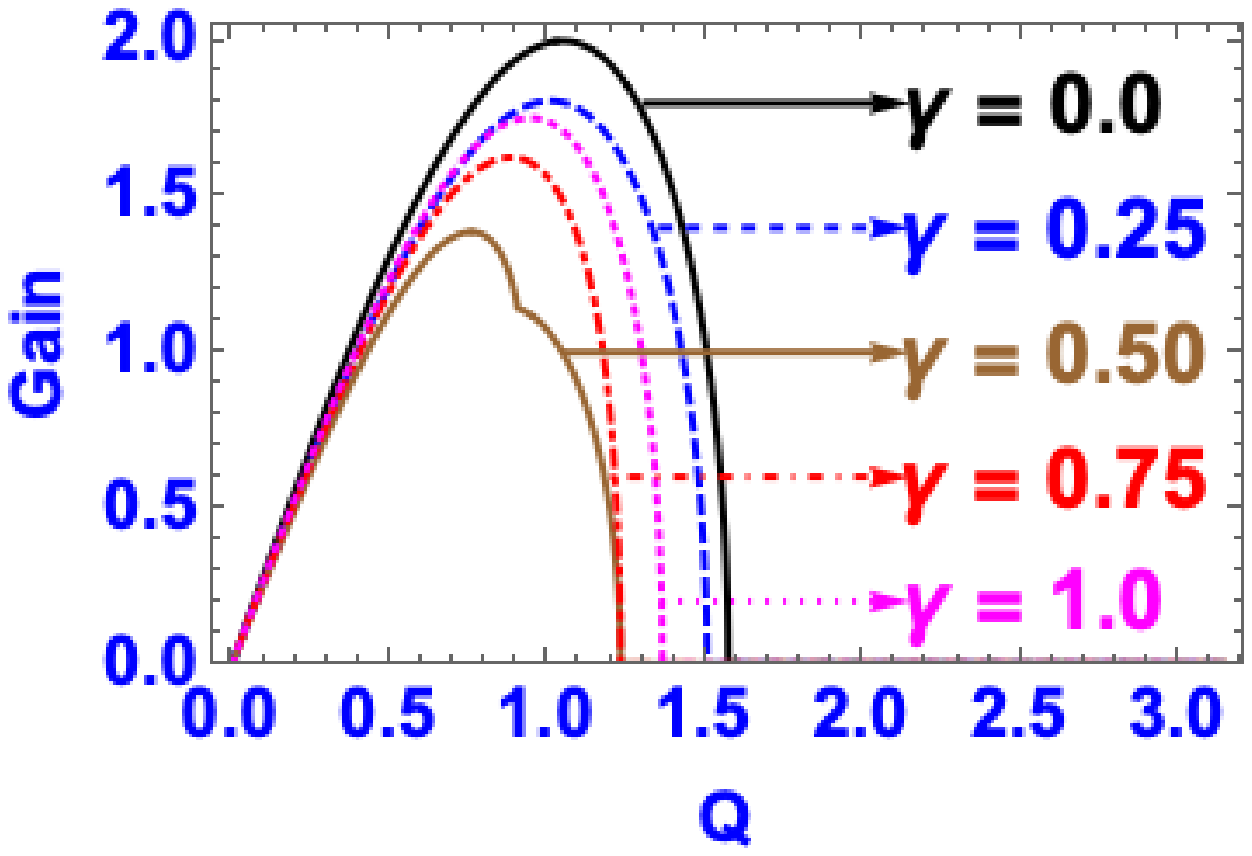}
\par
\includegraphics[width=.235\textwidth]{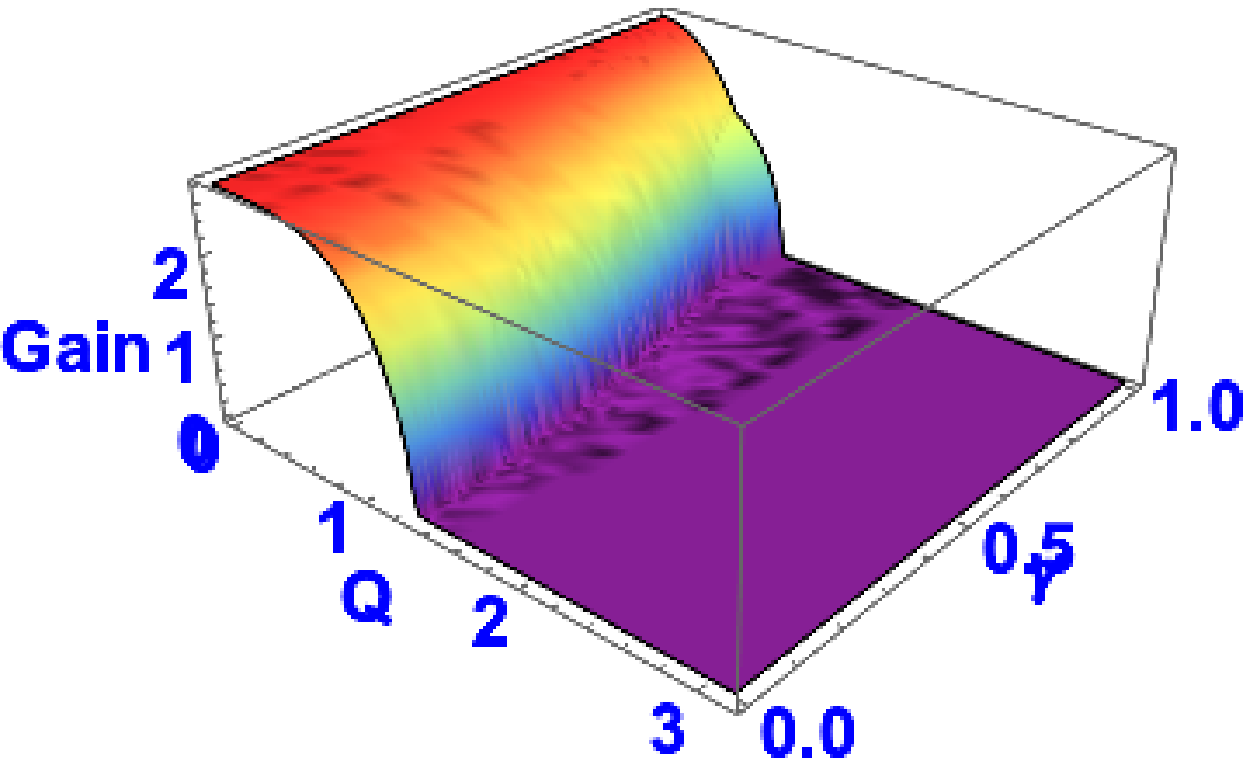} %
\includegraphics[width=.235\textwidth]{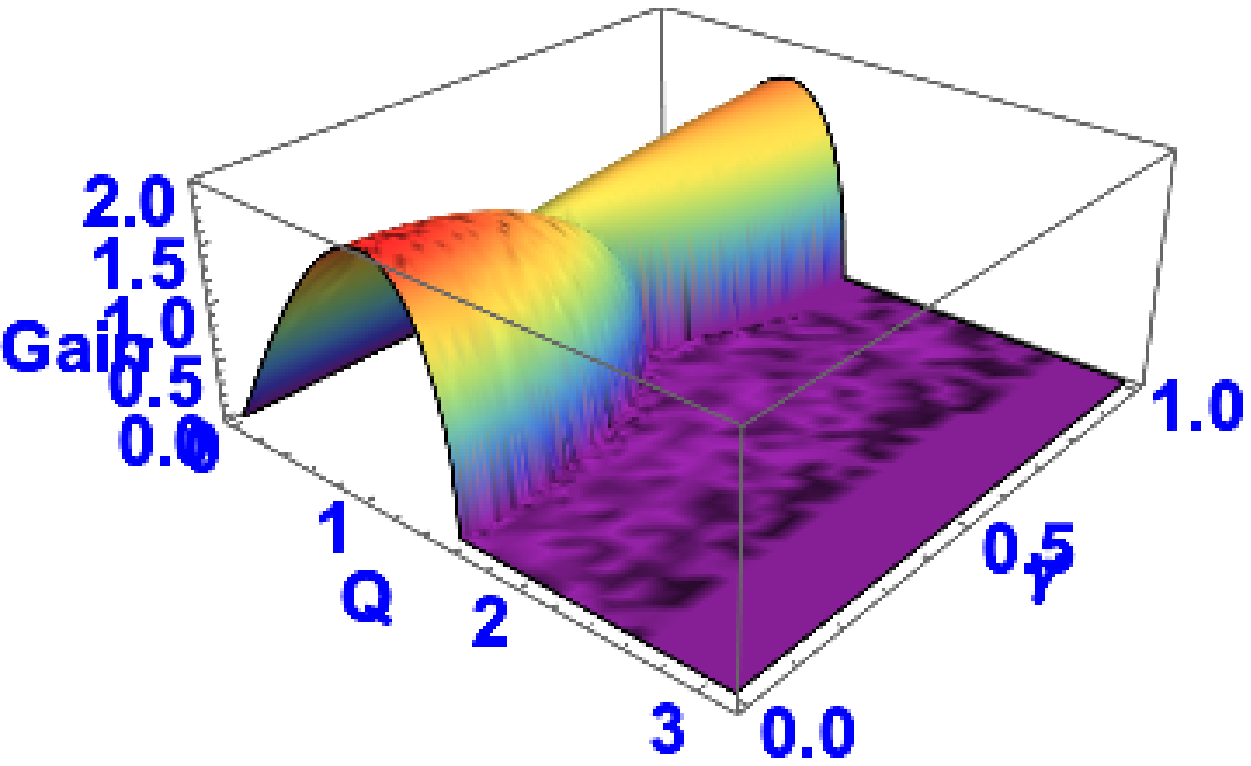}
\caption{(color online) Upper row: Two-dimensional (2D) plot showing the MI gain as a function of $Q$ and Lower row: Three-dimensional (3D) surface plot showing the MI gain as a function of $Q$ and $\gamma$.
Left and right columns correspond to $\delta=1$ and $\delta=0$, respectively. Remaining parameters are $g= 1$, $G=-1$ and $\Gamma=1$.}
\label{rf1}
\end{figure}
In Fig.\ref{rf1}, upper two panels show the MI gain as a function of $Q$ for five different values of SO coupling strengths, $\gamma=0.0, 0.25,0.50,0.75,$ and $1.0$. Left and right columns correspond to $\delta=1$ and $\delta=0$. As the strength of the SO coupling increases from $0$ to $0.5$, the MI gain decreases for both $\delta=1$ and $\delta=0$. As we increase the strength of SO coupling further, one observes no variation in the gain for $\delta=1$ . But, for $\delta=0$, the MI gain increases for $\gamma>0.5$. The three-dimensional (3D) surface plot clearly illustrates this behaviour in the lower panels. 


\subsubsection{Miscible / Immiscible condensates}
Next, we study the onset of  MI in Miscible / Immiscible condensates. 

3D surface plot shows the existence of MI domain as a function of $Q$ and $\gamma$ for upper row: partially miscible, ($g= 1$, $G=-1$), middle row: immiscible ($g= 1$, $G=-0.5$) and lower row: miscible ($g= 1$, $G=-1.5$) condensates.

\begin{figure}[htbp!]
\includegraphics[width=.235\textwidth]{caseUS3b_3d.eps} 
\includegraphics[width=.235\textwidth]{caseUS3a_3d.eps}
\par
\includegraphics[width=.235\textwidth]{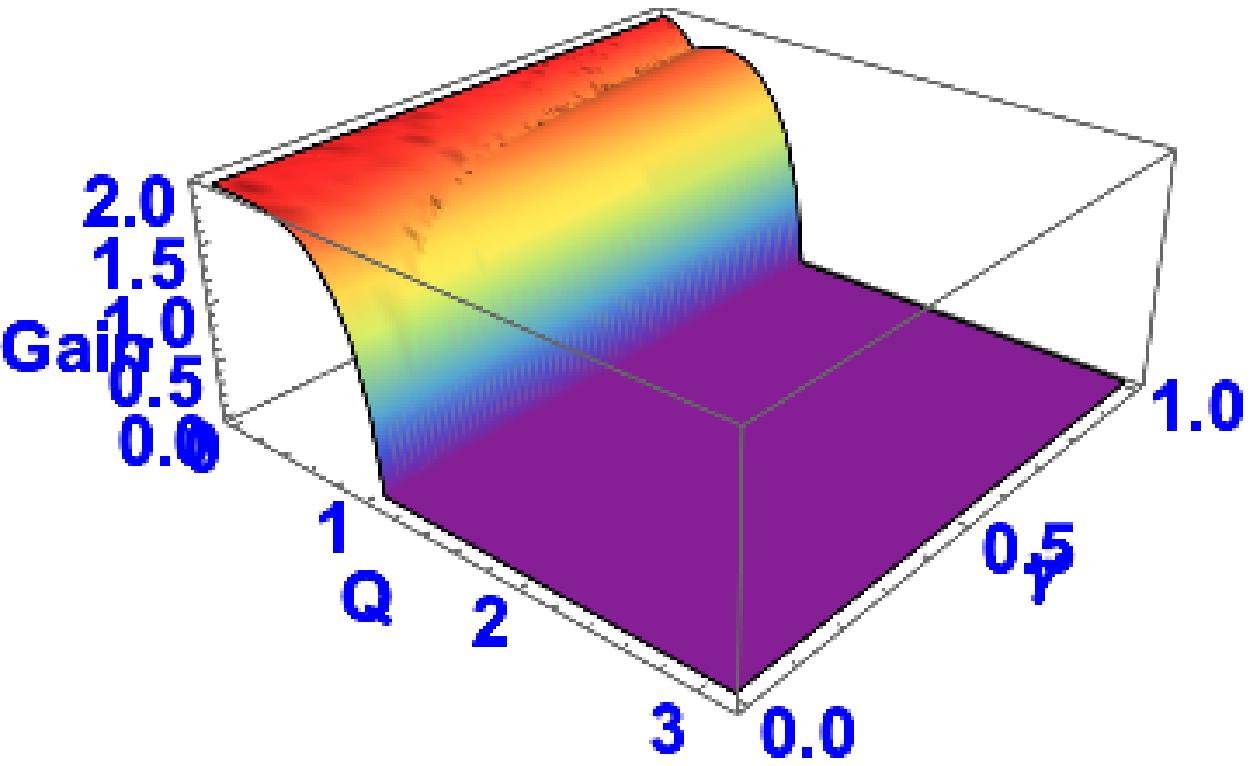}
\includegraphics[width=.235\textwidth]{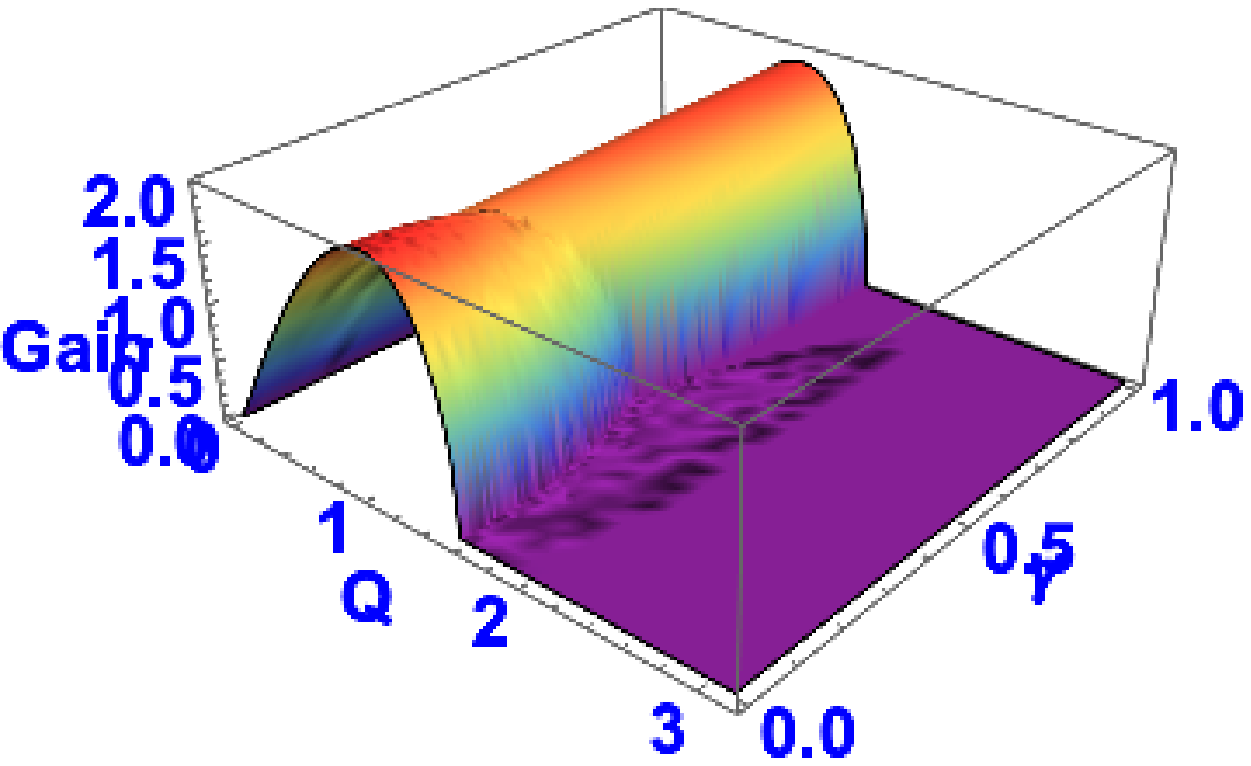}
\par 
\includegraphics[width=.235\textwidth]{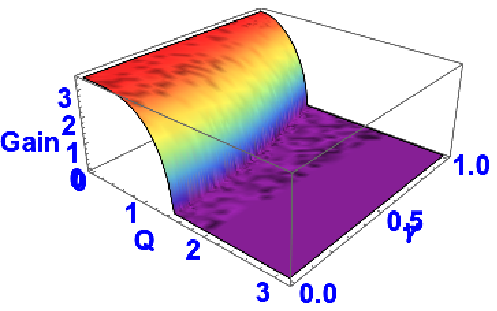}
\includegraphics[width=.235\textwidth]{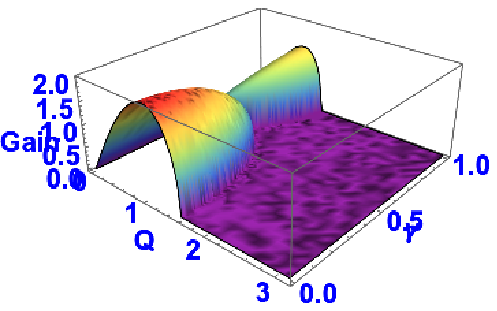}
\caption{(color online) 3D surface plot showing the MI region as a function of $Q$ and $\gamma$ for upper row: partially miscible, ($g= 1$, $G=-1$), middle row: immiscible ($g= 1$, $G=-0.5$) and lower row: miscible ($g= 1$, $G=-1.5$) conditions. Left and right columns correspond to $\delta=1$ and $\delta=0$, respectively.}
\label{rf3}
\end{figure}
\subsection{MI in Staggered Mode}
In this section, we study the onset of MI in a staggered mode. It is worth mentioning at this juncture that at $\gamma=0$, our results completely agree with the Ref.~\cite{ZRaptiMI}, and could reproduce the results of one of the cases (in the absence of three-body interaction) discussed in Ref.~\cite{Baizakov 2C MI}. These results are reproduced by our investigation.

Next, to study the impact  of SO coupling on MI i.e $\gamma\neq 0$, we first discuss the MI gain for $g>0$ and $G>0$. 
To facilitate our understanding and to make the investigation straightforward, we have reviewed the possible MI domains  in table 2. 
We have thoroughly brought out the MI domains in the absence/presence/ of hopping strength, SO coupling and Rabi coupling for different possible combinations of signs of intra- and inter-component interactions strength. 

\begin{table}[htbp!]
\caption {Review of the MI in discrete SOC-BECs at Staggered mode}
\label{table1}
\begin{center}
\begin{tabular}{|l|l|l|l|l|}
\hline 
\textbf{Hopping }& &\textbf{Rabi } & \textbf{Possible} &  \\
\textbf{Strength}&\textbf{SOC} &\textbf{coup.}& \textbf{Cases} & \textbf{Conjecture} \\
\hline \hline
& & & $g>0$, $g_{12}>0$  & \,\,\,\,Stable \\ 
& &  \,\,$\delta=0$ & $g>0$, $g_{12}<0$ & \,\,\,\,Stable \\ 
& & & $g<0$, $g_{12}>0$ & \,\,\,\,Unstable\\ 
&$\gamma=0$  & & $g<0$, $g_{12}<0$ & \,\,\,\,Unstable \\  \cline{3-5}
&\cite{ZRaptiMI} & &$g>0$, $g_{12}>0$  & \,\,\,\,Stable \\ 
& &  \,\,$\delta=1$ & $g>0$, $g_{12}<0$ & \,\,\,\,Unstable \\ 
& & & $g<0$, $g_{12}>0$ & \,\,\,\,Stable\\ 
\,\,\,\,\,$\Gamma=1$ & & & $g<0$, $g_{12}<0$ & \,\,\,\,Unstable \\  
\cline{2-5}
& & &$g>0$, $g_{12}>0$  & \,\,\,\,Stable \\ 
& &  \,\,$\delta=0$ & $g>0$, $g_{12}<0$ & \,\,\,\,Stable \\ 
& & & $g<0$, $g_{12}>0$ & \,\,\,\,Unstable\\ 
&$\gamma=1$  & & $g<0$, $g_{12}<0$ & \,\,\,\,Unstable \\  \cline{3-5}
& & &$g>0$, $g_{12}>0$  & \,\,  \\ 
& &  \,\,$\delta=1$ & $g>0$, $g_{12}<0$ & \,\,\,\,Stable \\ 
& & & $g<0$, $g_{12}>0$ & \,\, \\ 
& & & $g<0$, $g_{12}<0$ & \,\,  \\  
\cline{1-5}
& & &$g>0$, $g_{12}>0$  & \,\, \\ 
& &  \,\,$\delta=0$ & $g>0$, $g_{12}<0$ & \,\,\,\,Unstable \\ 
& & & $g<0$, $g_{12}>0$ & \,\, \\ 
\,\,\,\,\,$\Gamma=0$ &$\gamma=1$  & & $g<0$, $g_{12}<0$ & \,\,  \\  \cline{3-5}
& & &$g>0$, $g_{12}>0$  & \,\,\,\,Unstable \\ 
& &  \,\,$\delta=1$ & $g>0$, $g_{12}<0$ & \,\,\,\,Unstable\\ 
& & & $g<0$, $g_{12}>0$ & \,\,\,\,Stable\\ 
& & & $g<0$, $g_{12}<0$ & \,\,\,\,Stable \\  
\hline
\end{tabular}
\end{center}
\end{table}

\begin{figure}[htbp!]
\includegraphics[width=.235\textwidth]{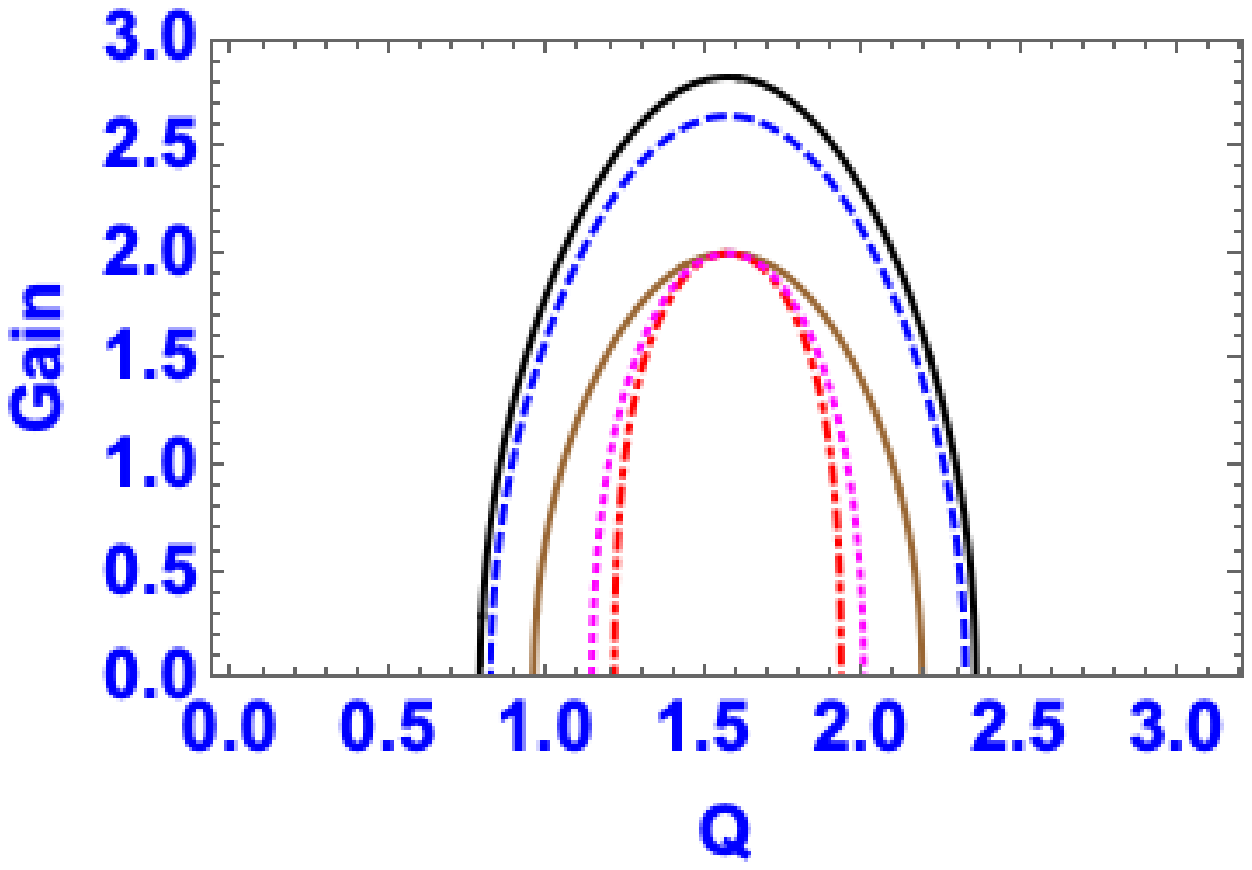} %
\includegraphics[width=.235\textwidth]{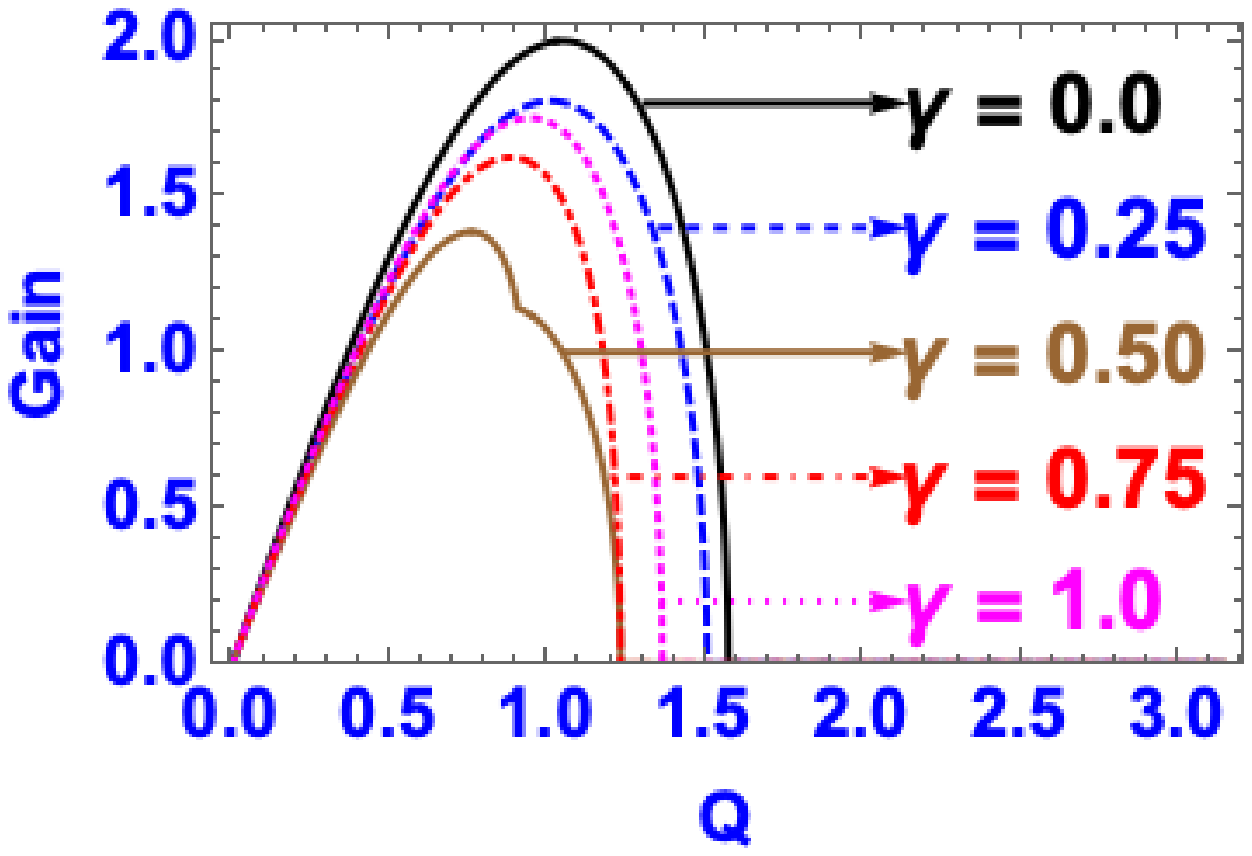}
\par
\includegraphics[width=.235\textwidth]{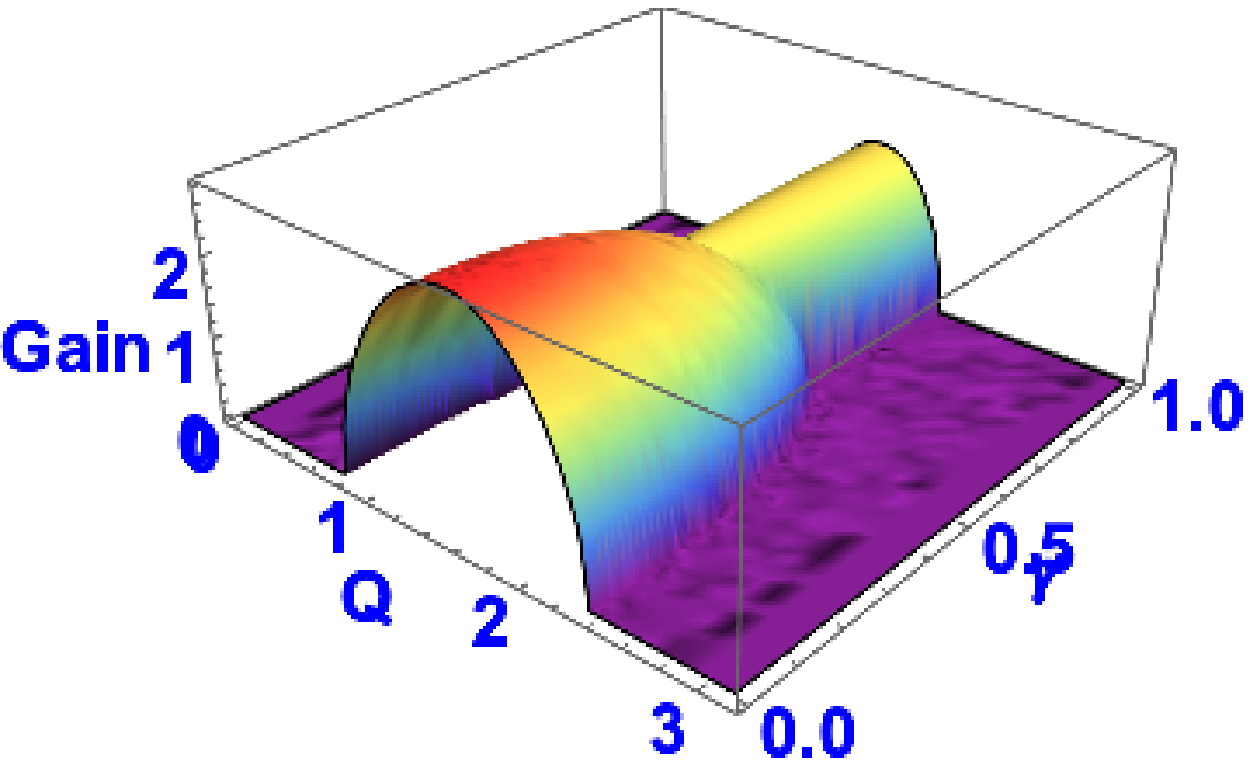} %
\includegraphics[width=.235\textwidth]{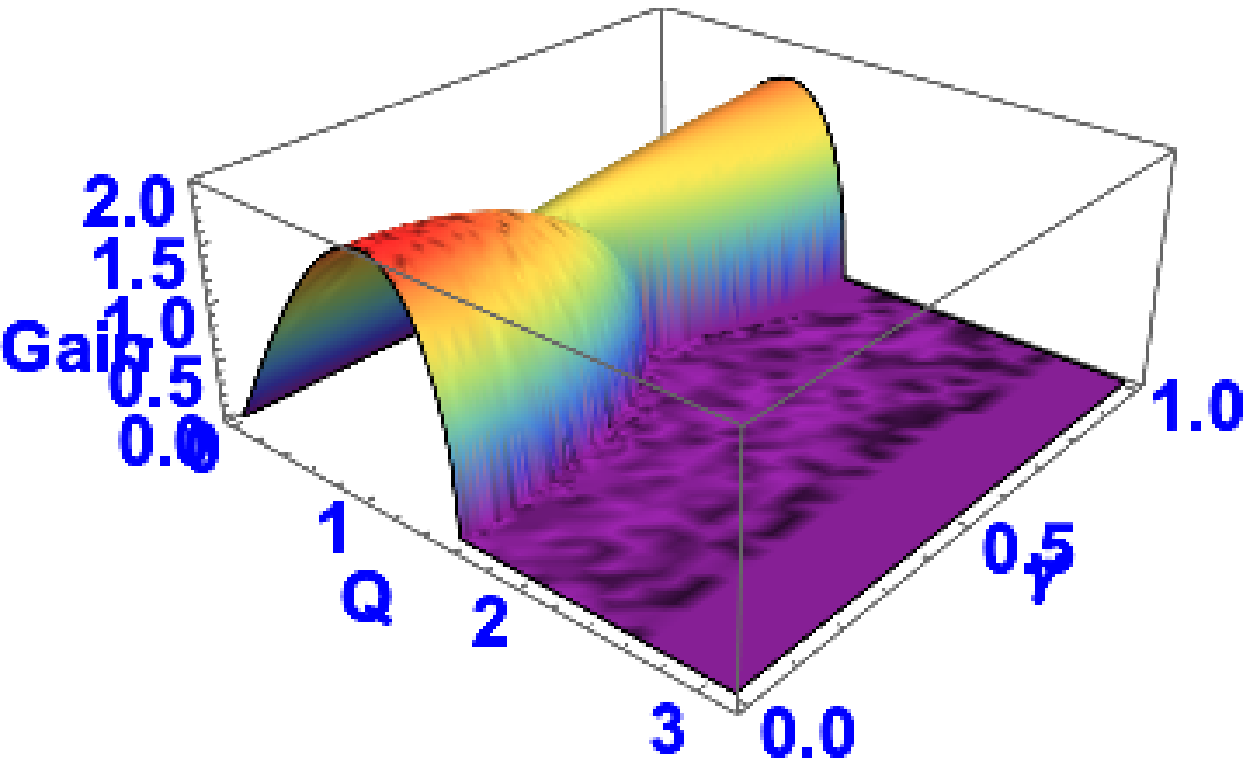}
\caption{(color online) Left and right columns correspond to $\delta=1$ and $\delta=0$, respectively. The remaining  parameters are $g= 1$, $G=1$ and $\Gamma=1$.}
\label{f3}
\end{figure}

In Fig.\ref{f3}, upper two panels show the MI gain as a function of $Q$ for five different values of SO coupling strengths, $\gamma=0.0, 0.25,0.50,0.75,$ and $1.0$. Left and right columns correspond to $\delta=1$ and $\delta=0$. As the strength of the SO coupling increases from $0$ to $0.5$, the MI gain decreases for both $\delta=1$ and $\delta=0$ . As we increase the strength of SO coupling further, one observes no variation in the gain for $\delta=1$ . But, for $\delta=0$ , the MI gain increases for $\gamma>0.5$. The three-dimensional (3D) surface plot clearly illustrates this behaviour in the lower panels. 

\begin{figure}[htbp!]
\includegraphics[width=.235\textwidth]{case1b_2d.eps} %
\includegraphics[width=.235\textwidth]{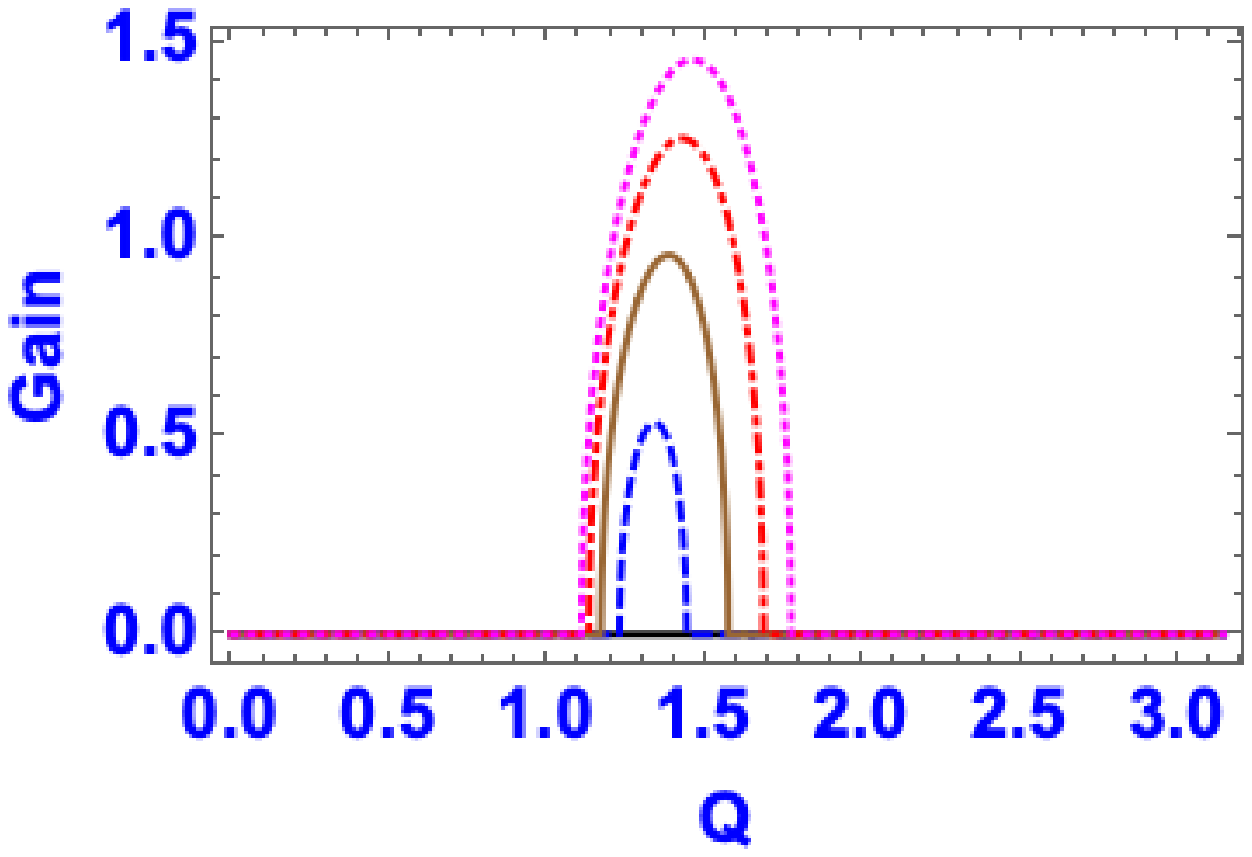}
\par
\includegraphics[width=.235\textwidth]{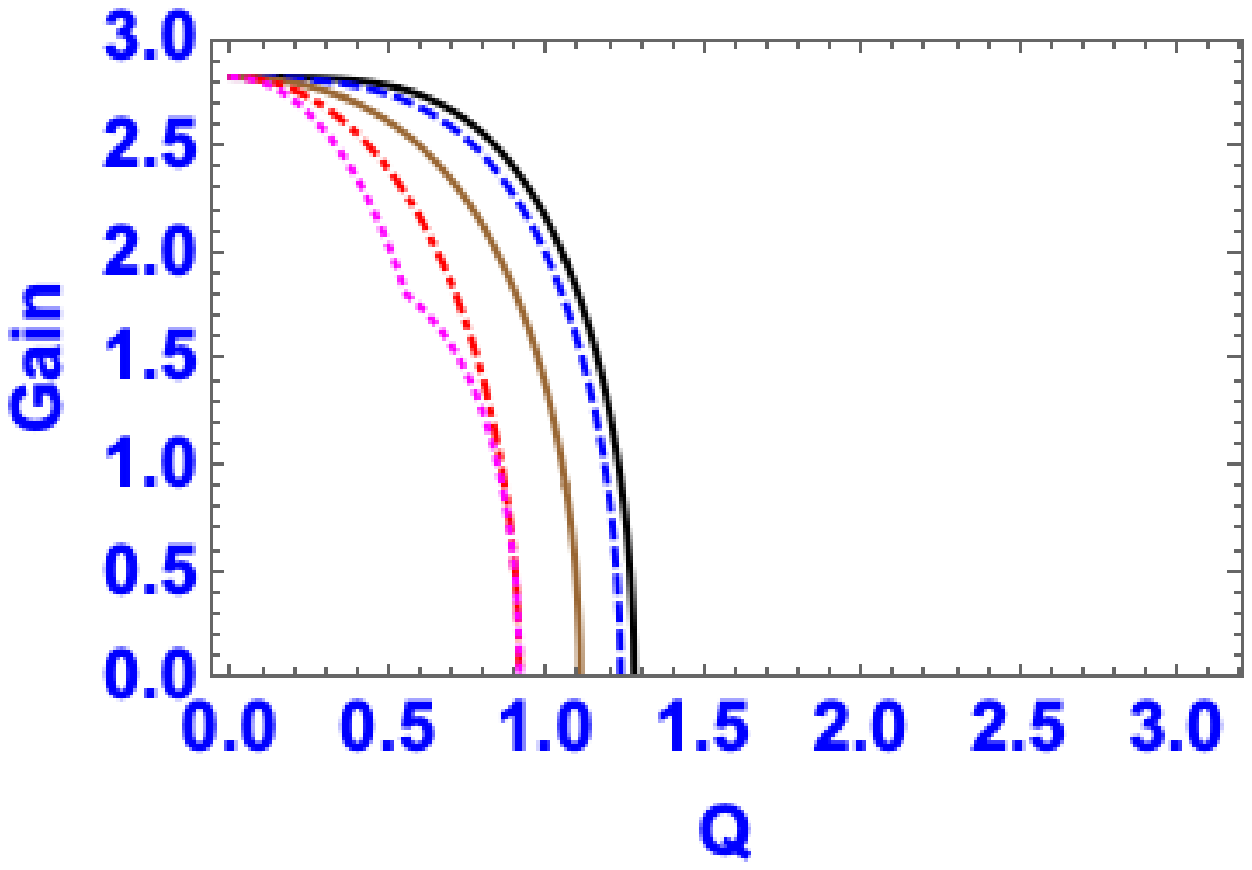} %
\includegraphics[width=.235\textwidth]{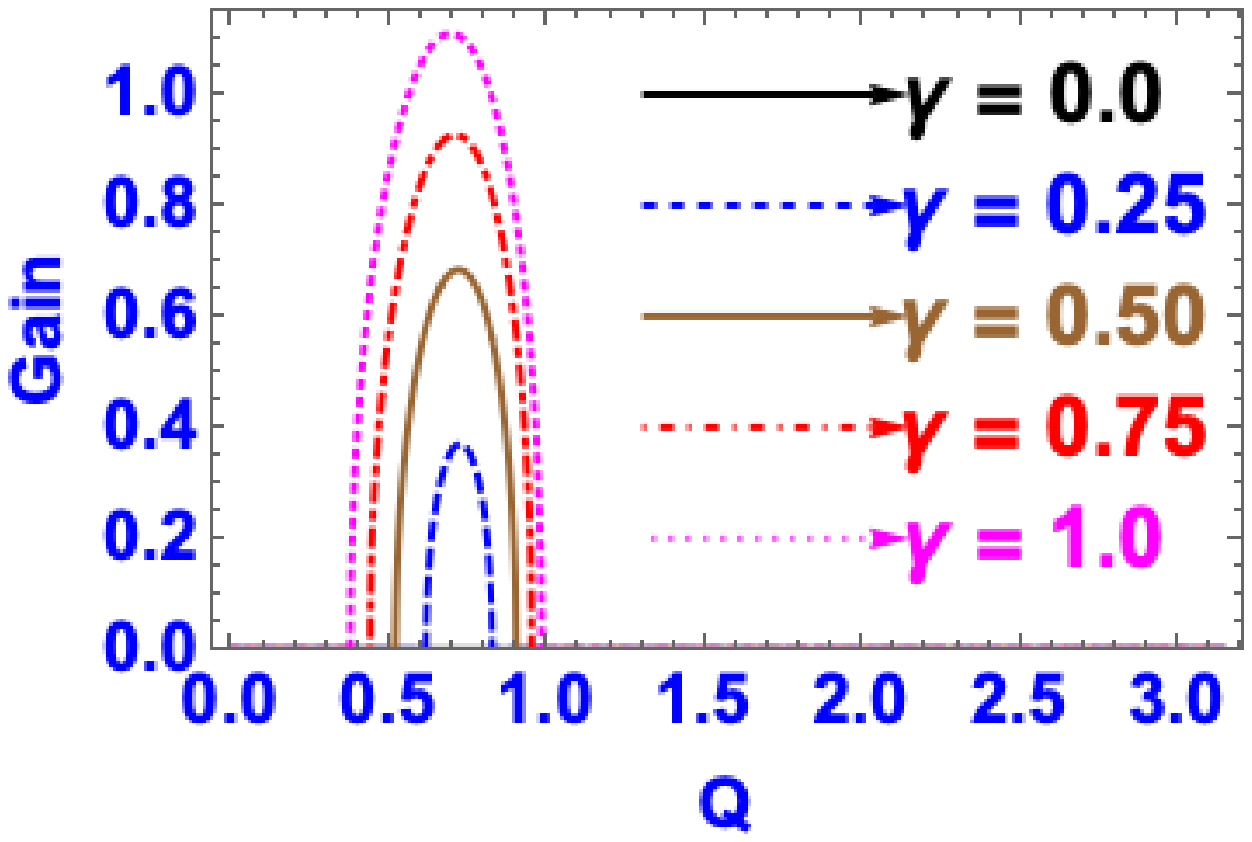}
\caption{(color online) MI gain as a function of $Q$ for (top left to bottom right) (i) $g= 1$, $G=1$, (ii) $g= 1$, $G=-1$,(iii) $g= -1$, $G=1$ and (iv) $g=-1$, $G=-1$. Remaining parameters are $\delta=1$ and $\Gamma=1$.}
\label{f5}
\end{figure}
\begin{figure}[htbp!]
\includegraphics[width=.235\textwidth]{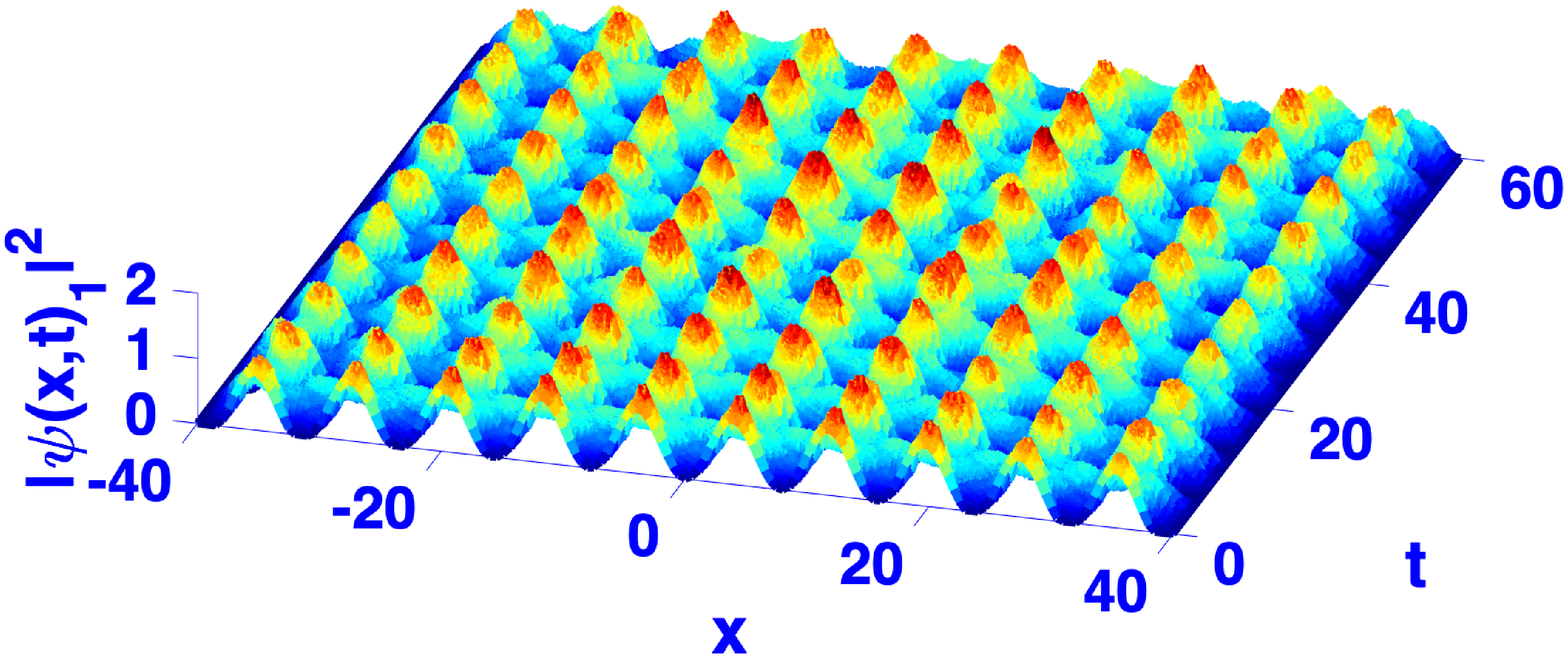} %
\includegraphics[width=.235\textwidth]{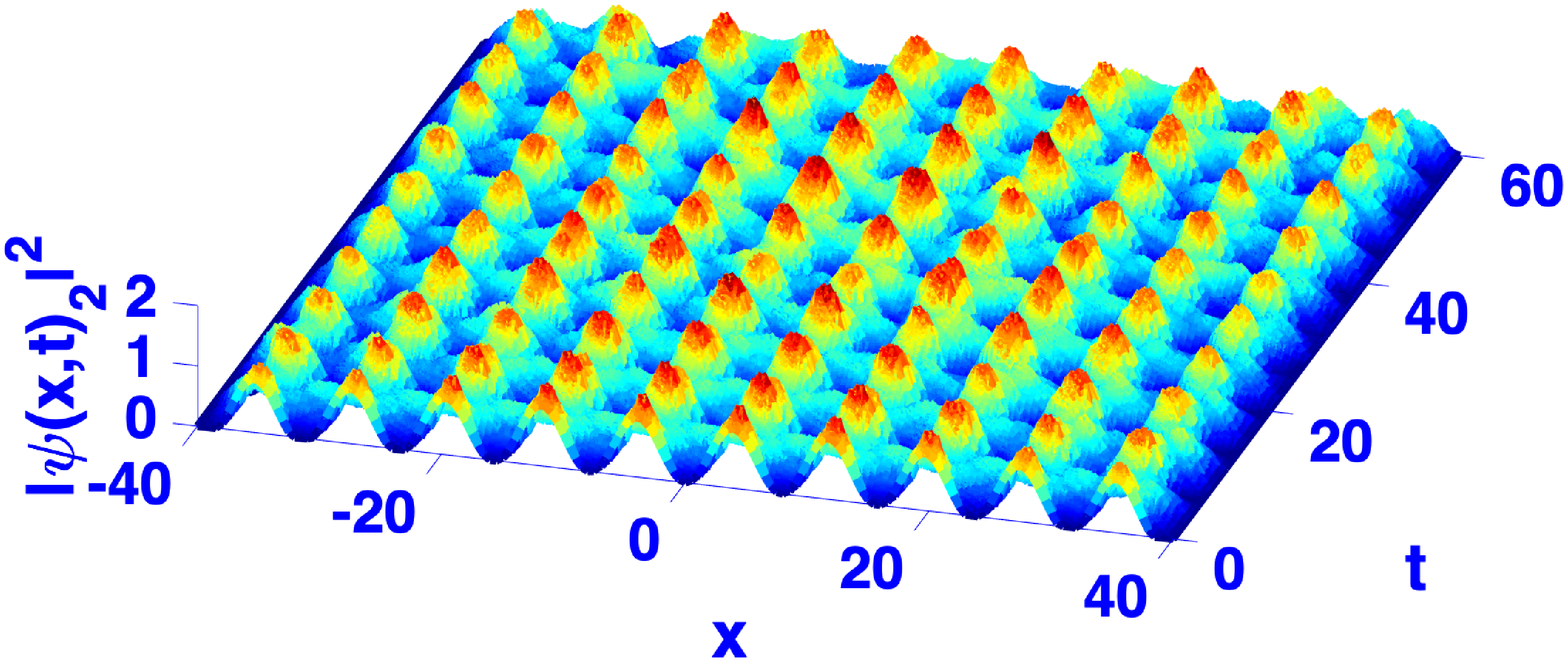}
\par
\includegraphics[width=.235\textwidth]{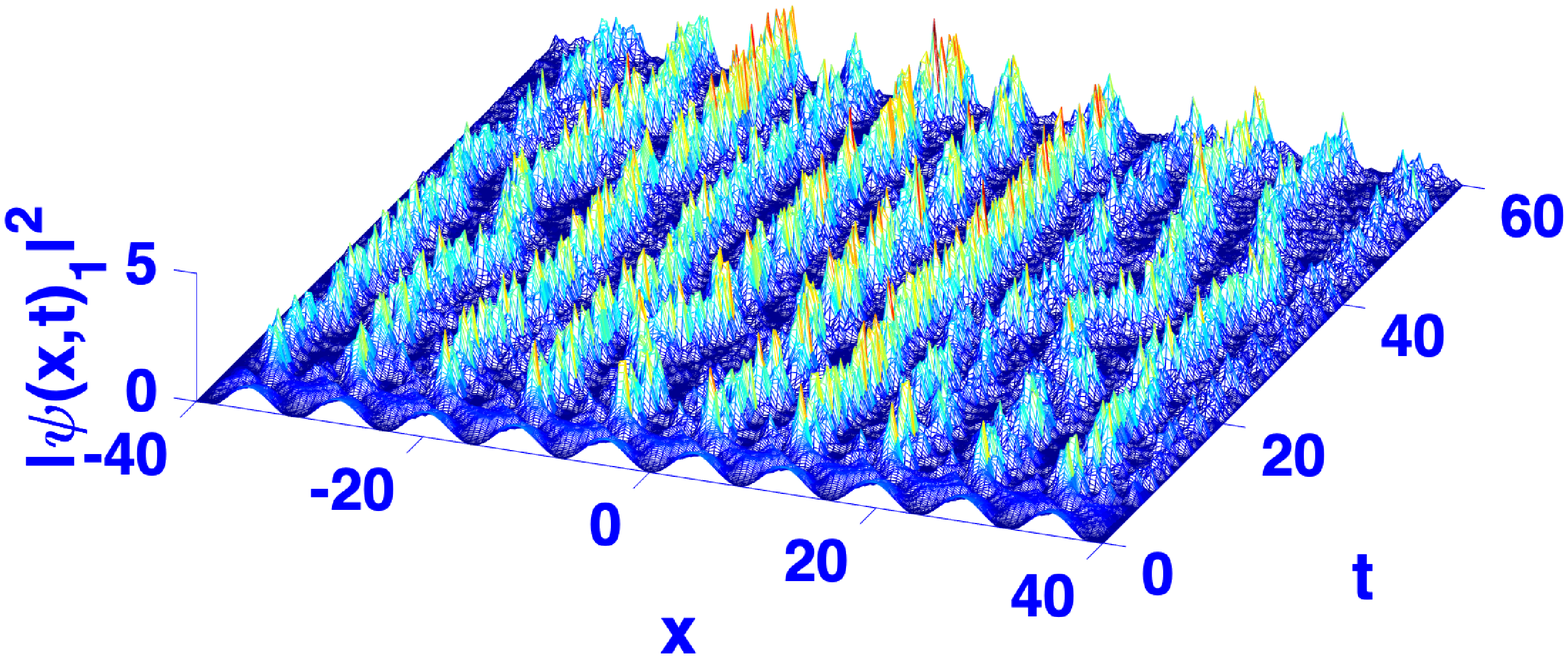} %
\includegraphics[width=.235\textwidth]{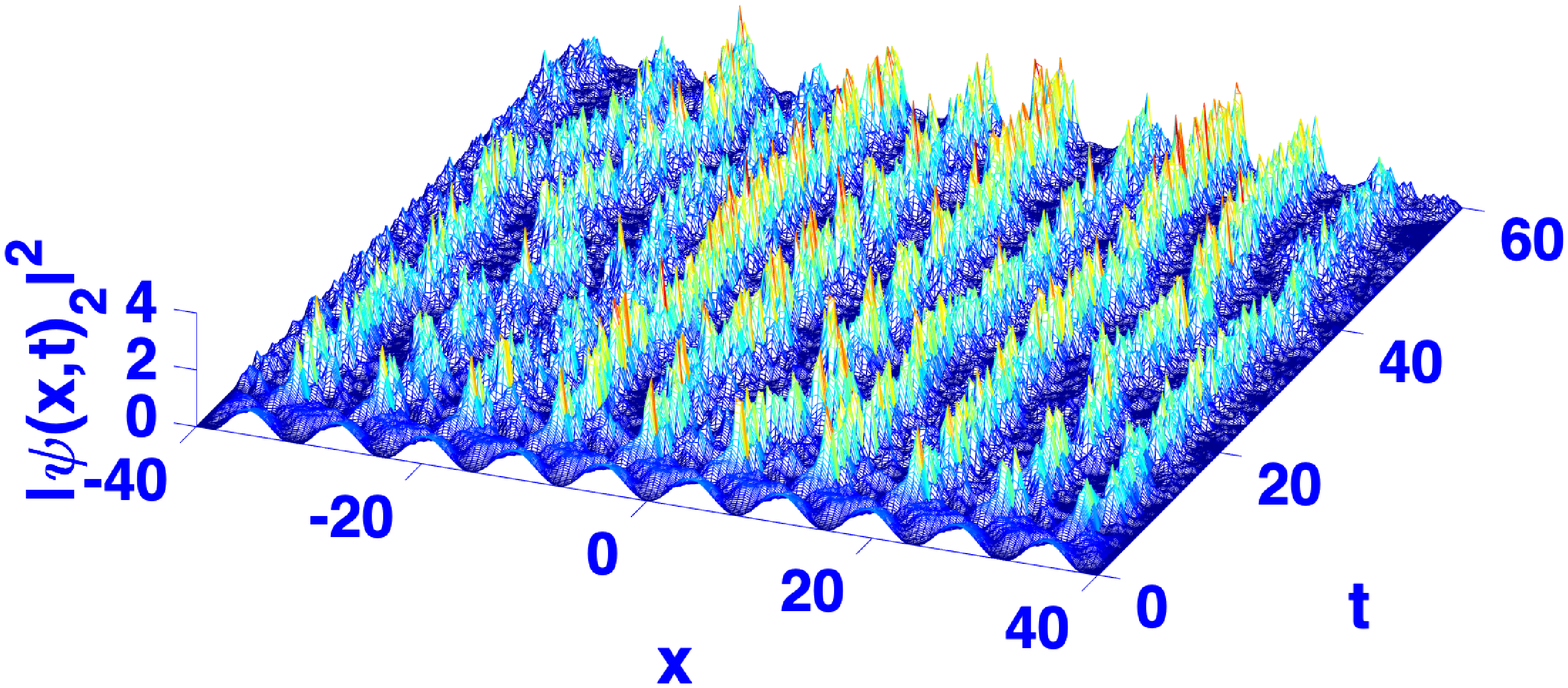}
\par
\includegraphics[width=.49\textwidth]{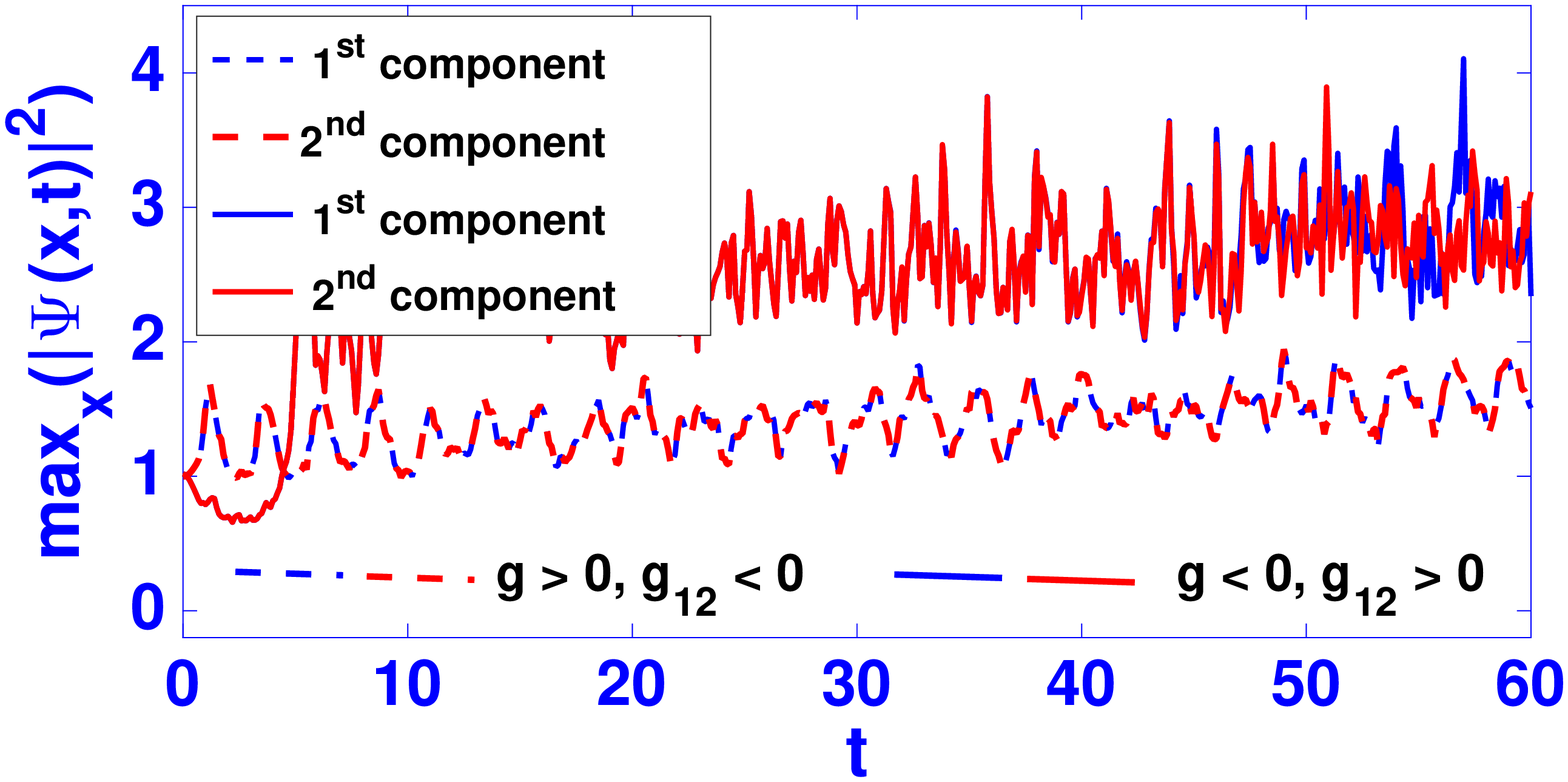}
\caption{(Upper and middle) The space-time evolution of the density in both condensates (left and right) for  MS and MU, $g>0$, $g_{12}<0$ and $g<0$, $g_{12}>0$, respectively. Lower panel shows the comparative time evolution of the maximum density of MU and MS for the same parameters corresponding to $2^{nd}$ and $3^{rd}$ cases in Table 1.)}
\label{fnu1}
\end{figure}
Next, we discuss the MI domains for other three different possible combinations of intra- and inter-species fragmentation, i.e $g$ and $G$. When the inter-component interaction is repulsive ($G>0$), the MI gain decreases when one increases the strength of SO coupling and this occurs irrespective of the signs of intra-component interaction. Also, when $G<0$, the MI gain increases when we  increase $\gamma$. Further, one does not observe MI when $\gamma=0$ for both $G<0$  as shown in Fig.\ref{f5}.

Figure~5 shows the modulationally stable (MS) and modulationally unstable (MU) behaviour for $2^{nd}$ and $3^{rd}$ cases corresponding to $g>0$, $g_{12}<0$ and $g<0$, $g_{12}>0$, in Table 1. Upper and middle panels show the 3D space-time evolution of the density for both condensates (left and right) and  for MS and MU cases, respectively. 
In the upper two panels, during the time evolution of the wave in condensates, the wave amplitudes (periodicity) exhibit small oscillations and remain close to their initial values demonstrating the stability of the system. On the contrary, in the lower two panels, the wave amplitude in the condensates grows exponentially due to their instability. 
The comparative evolution of the wave amplitudes in both cases is plotted in lower panel. The maximum density shows mild oscillations nearly around its initial value 1 for the MS case in both condensates (dashed blue and dash-dotted red lines). Blue and red solid lines show the exponential growth of the maximum densities for the MU mode in both the condensates.
\color{black}

\begin{figure}[htbp!]
\includegraphics[width=.235\textwidth]{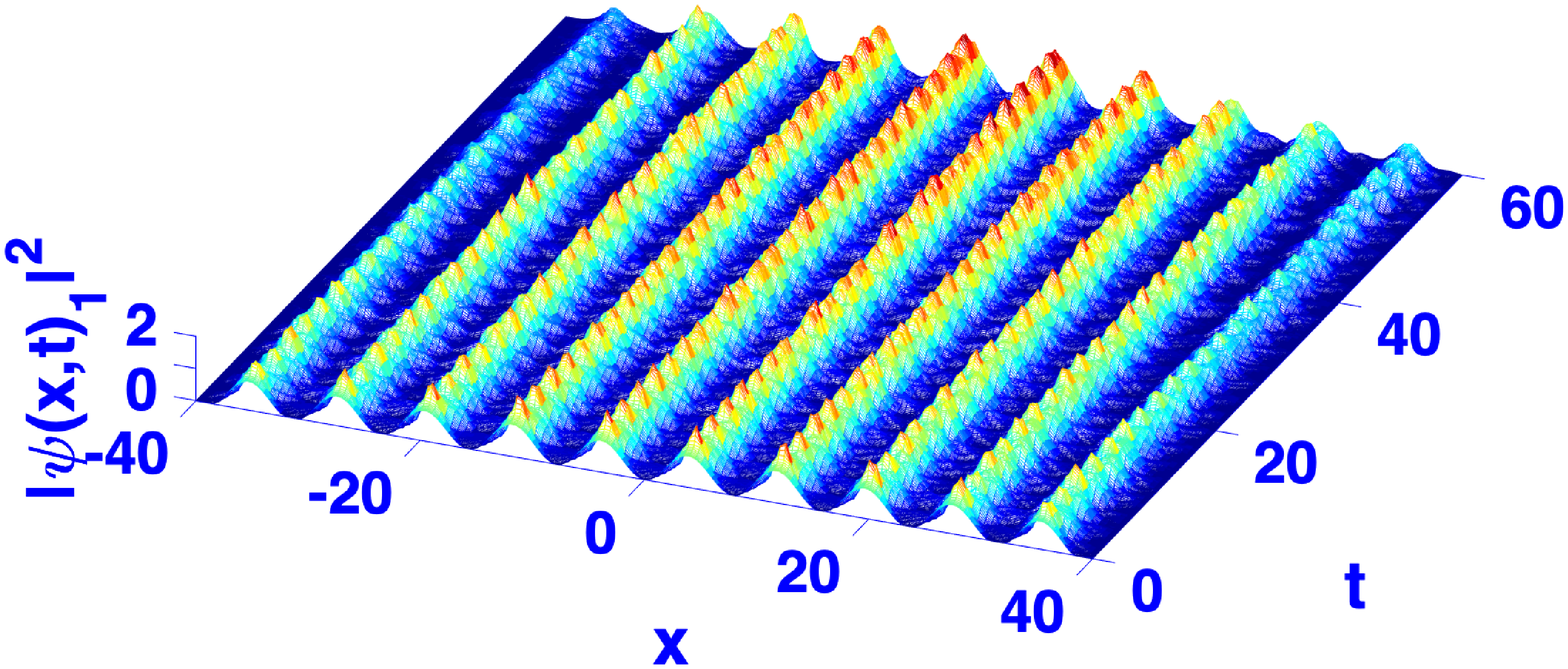} %
\includegraphics[width=.235\textwidth]{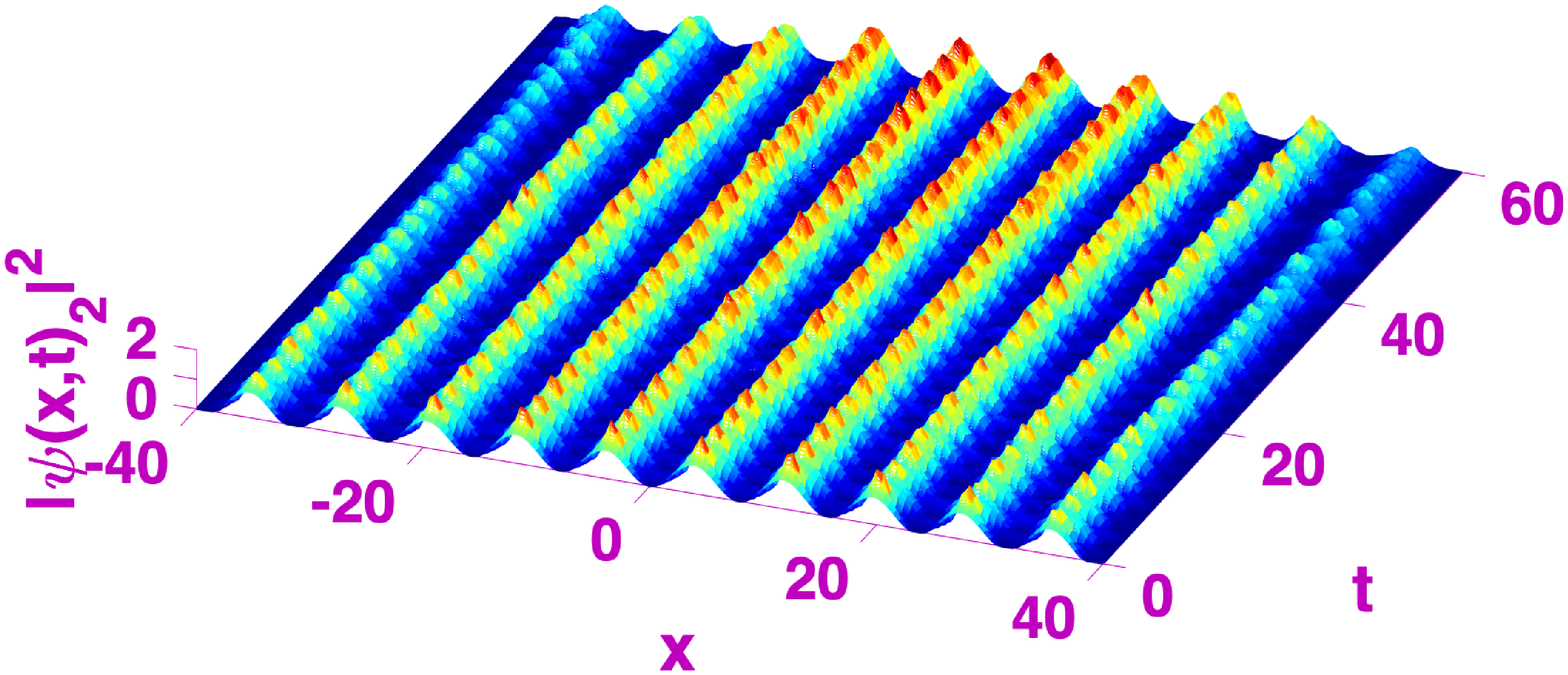}
\includegraphics[width=.235\textwidth]{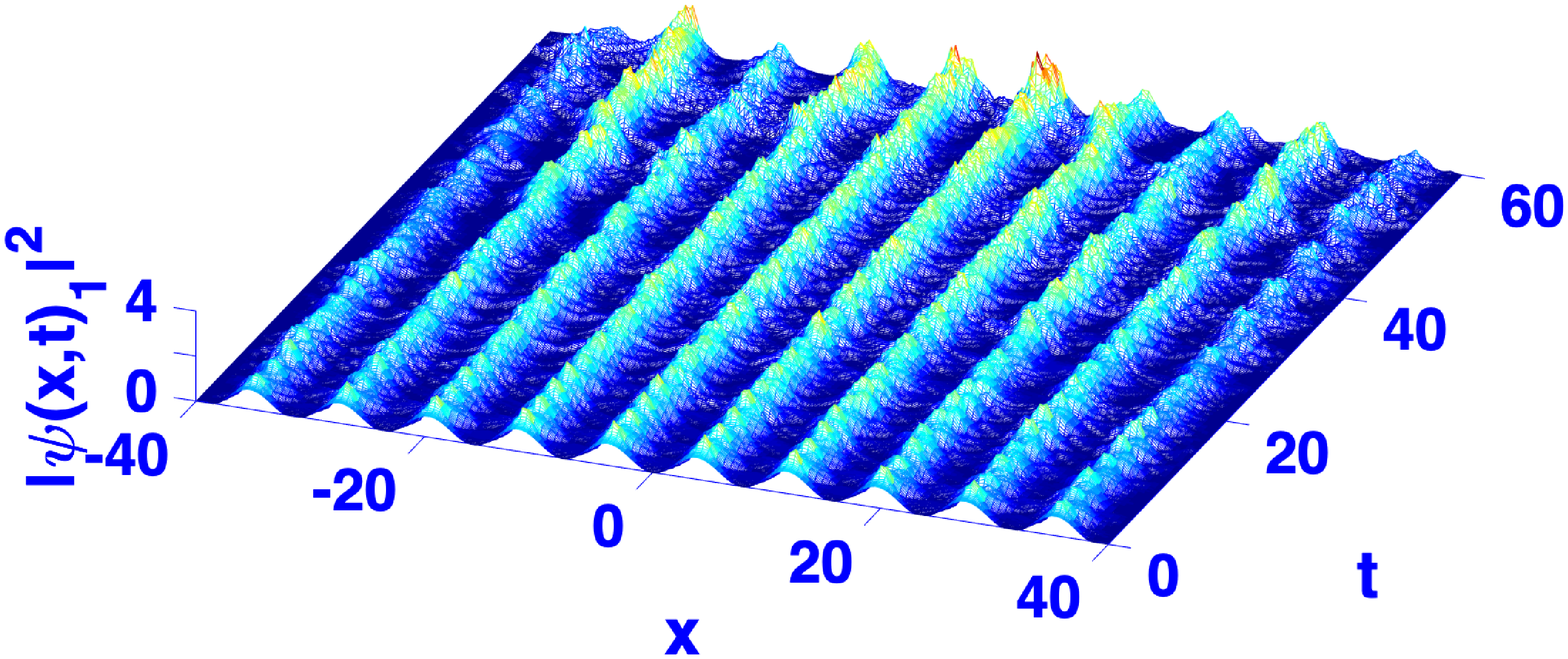} %
\includegraphics[width=.235\textwidth]{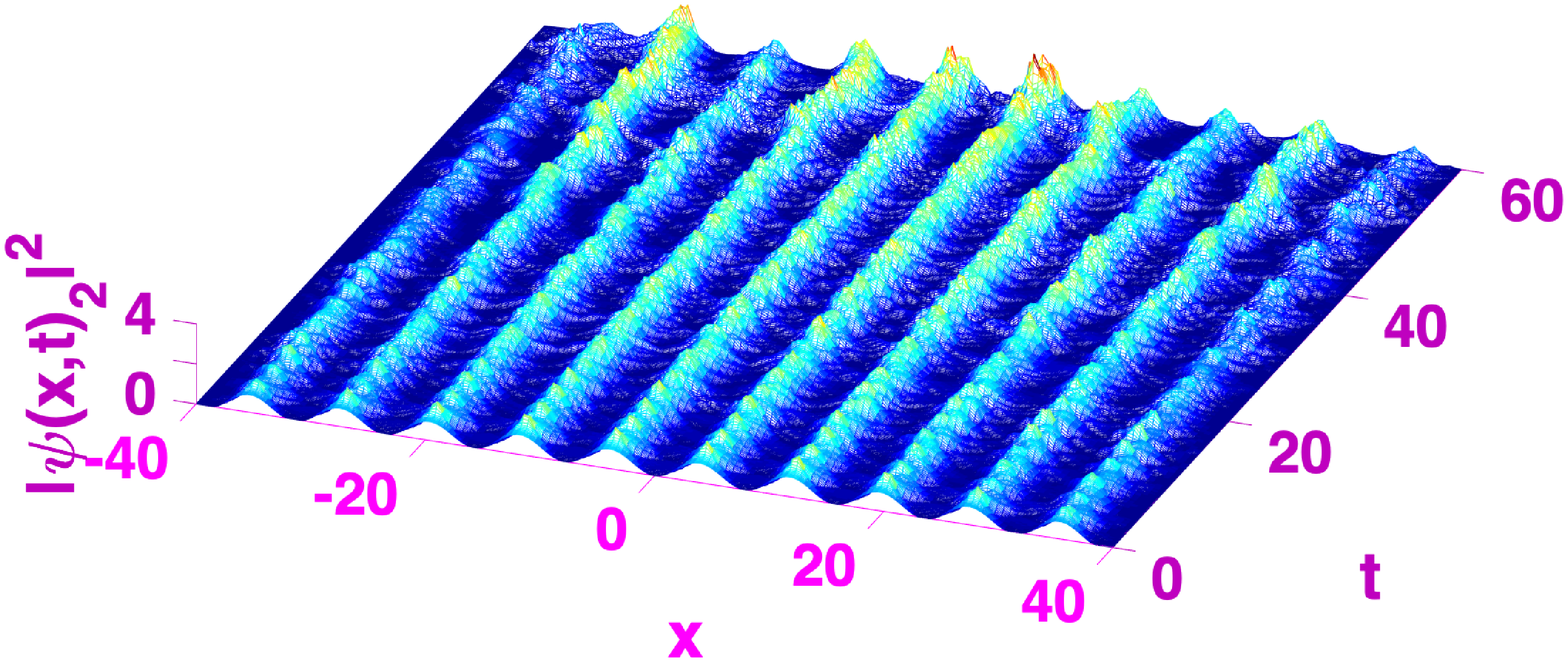}
\includegraphics[width=.235\textwidth]{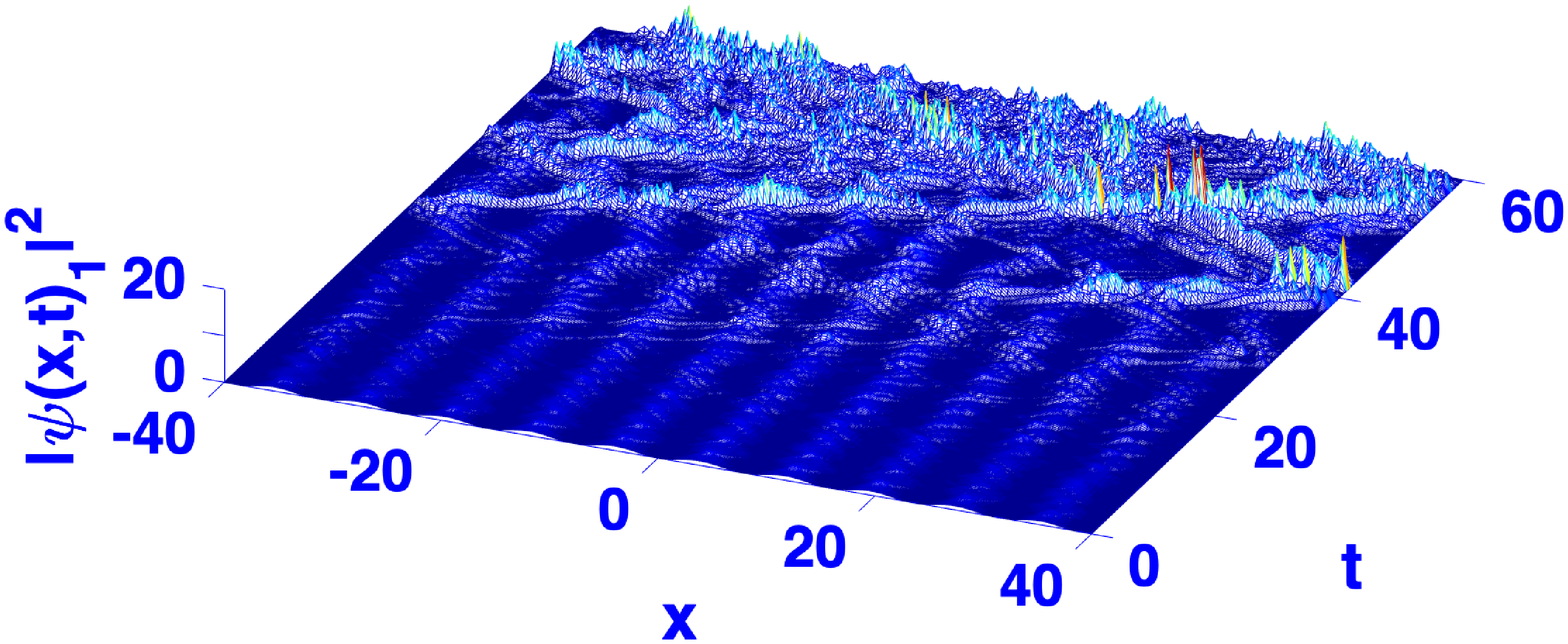} %
\includegraphics[width=.235\textwidth]{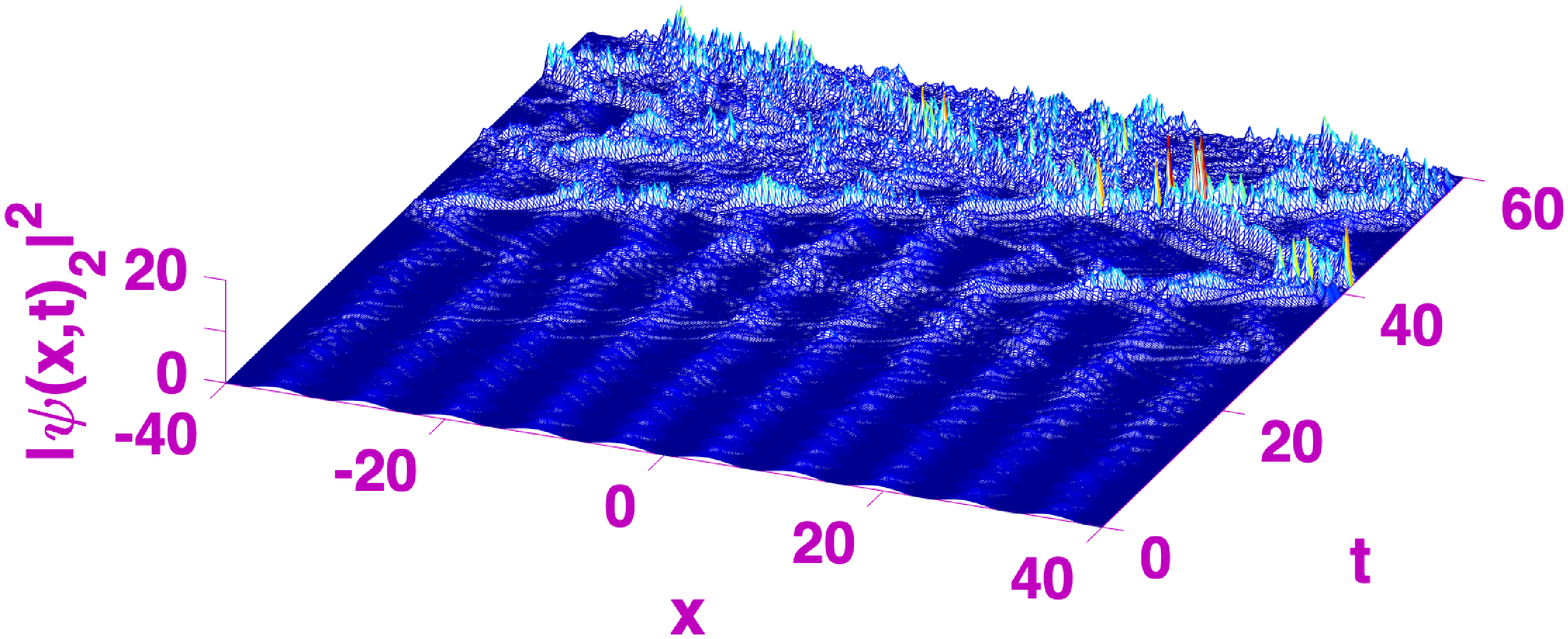}
\includegraphics[width=.235\textwidth]{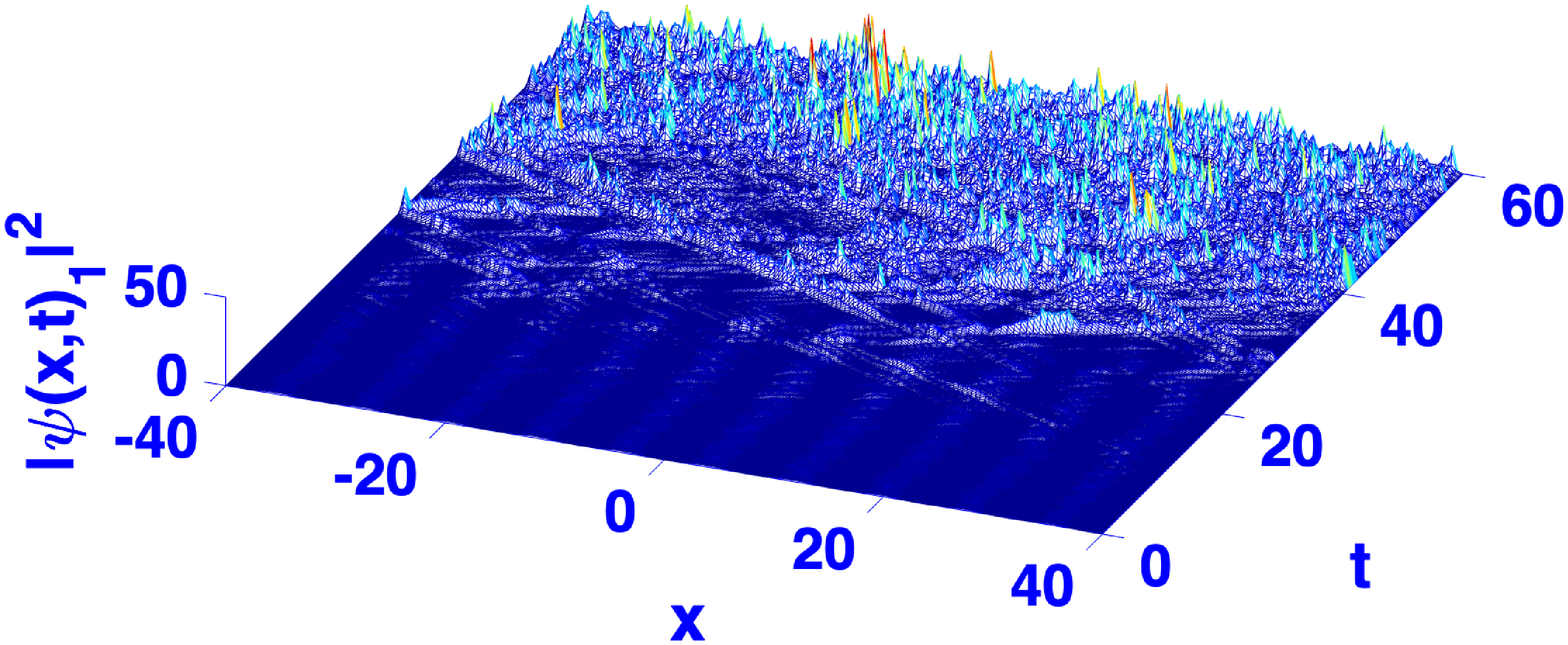} %
\includegraphics[width=.235\textwidth]{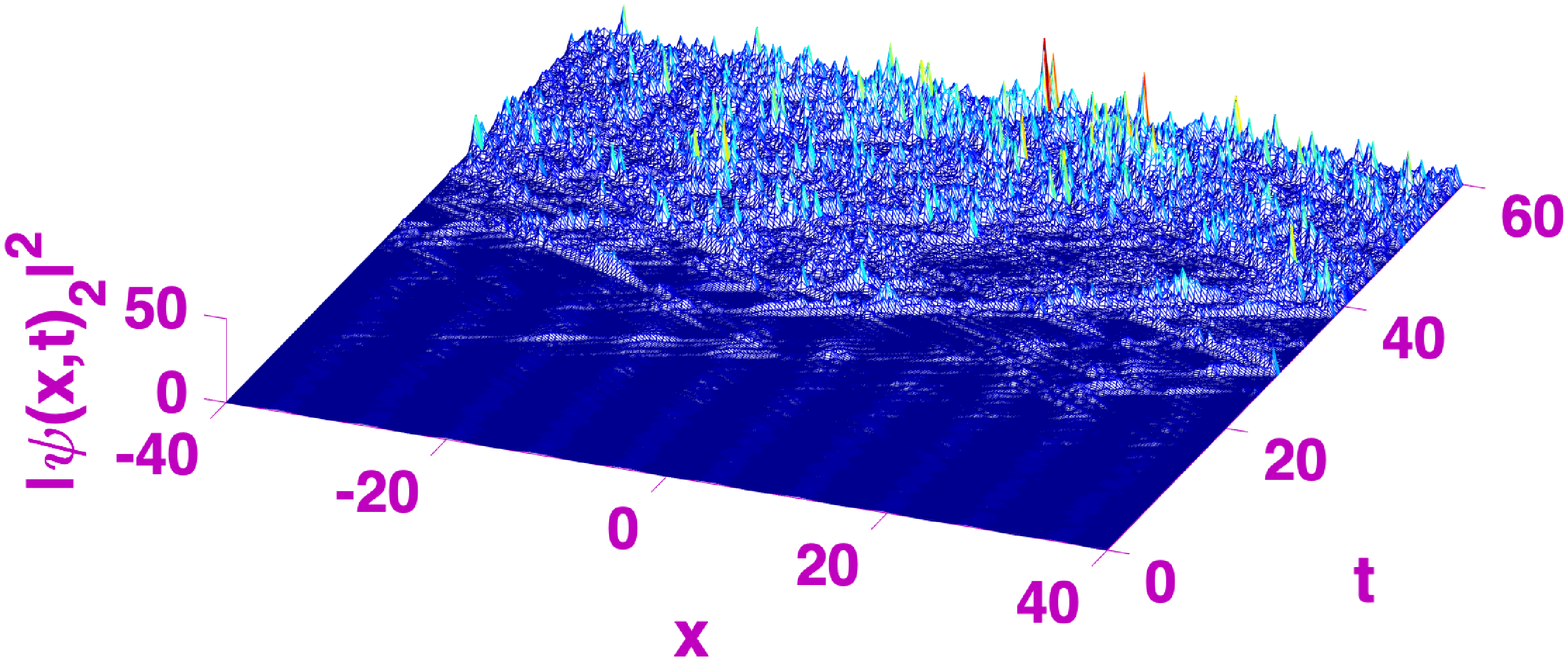}
\includegraphics[width=.235\textwidth]{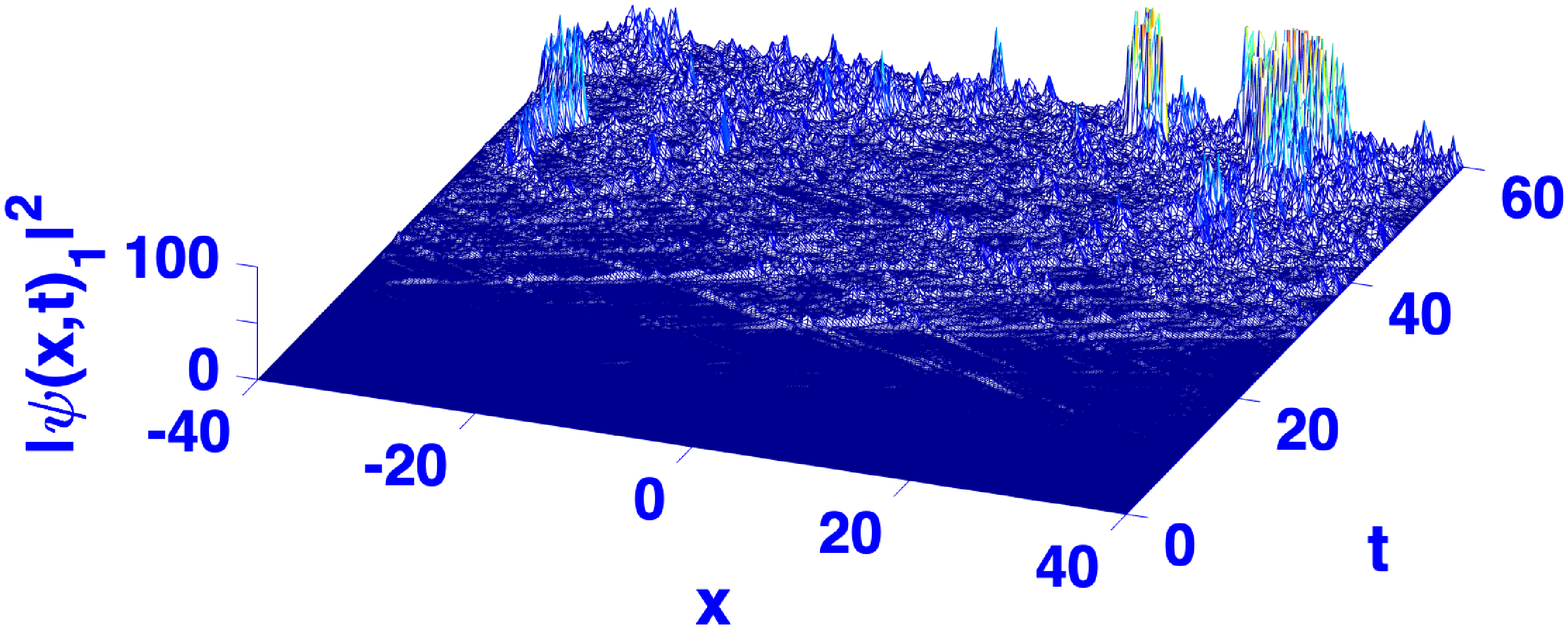} %
\includegraphics[width=.235\textwidth]{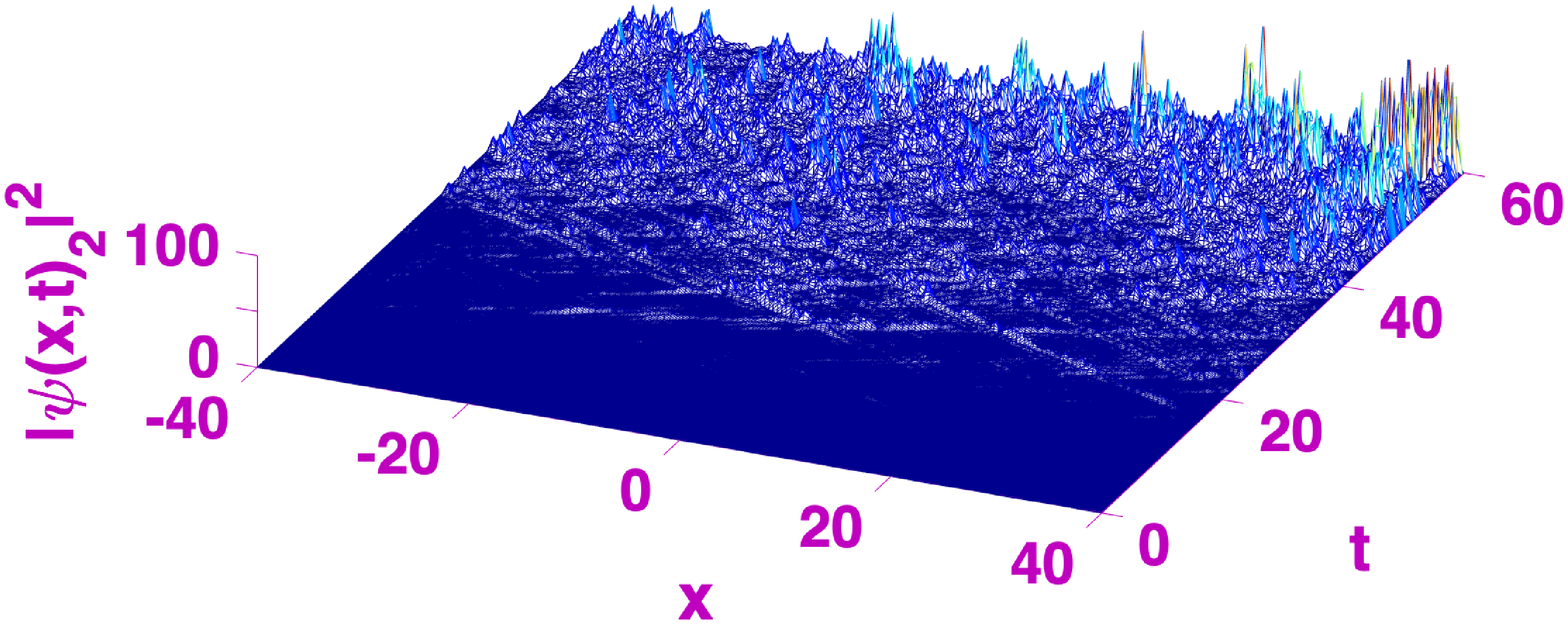}
\caption{The space-time evolution of the density in both the condensates (left and right). Panels from upper to lower show the density with increasing strengths of the SOC, $\gamma=0.1$, $\gamma=0.2$, $\gamma=0.3$, $\gamma=0.4$ and $\gamma=0.5$, respectively. Remaining parameters are $g=1.0$, $g_{12}=-1.0$ and $\delta=1.0$ with the other parameters picked from Fig.\ref{f5} (ii).}
\label{fnu2}
\end{figure}

Figure~6 shows the 3D space-time evolution of the density in both the condensates for the unstable modes. The instability increases on  increasing the strengths of the SOC. 
The 3D space-time evolution of the density remains stable upto time interval $t=60$ with small disturbances in both the condensates (left and right) for $\gamma=0.1$. However, when we increase the strength of the SOC to $\gamma=0.3$, the 3D space-time evolution of the densities starts to collapse around time $t=20$. Also, if we increase the strength of the SOC to $\gamma=0.5$ further, the maximum densities increase exponentially with large oscillations.

\subsubsection{Miscible / Immiscible condensates}
Next, we study the MI gain in Miscible / Immiscible condensates. 

Figure~7 displays the MI region in 3D surface plot as a function of $Q$ and $\gamma$ for  partially miscible ($g= 1$, $G=1$) condensates in upper row, miscible ($g= 1$, $G=0.5$) case in middle row and immiscible ($g= 1$, $G=1.5$) case in lower row.
%
\begin{figure}[htbp!]
\includegraphics[width=.45\textwidth]{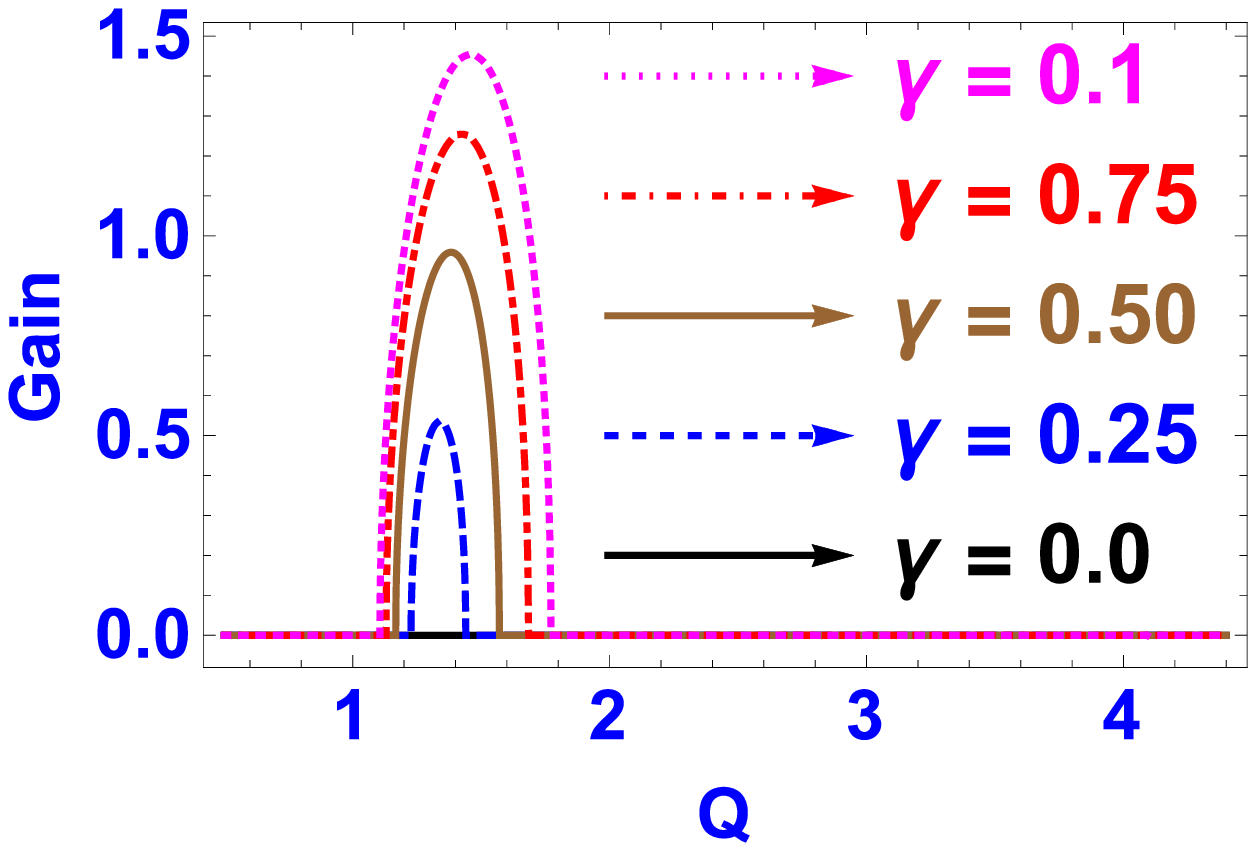} %
\par
\includegraphics[width=.235\textwidth]{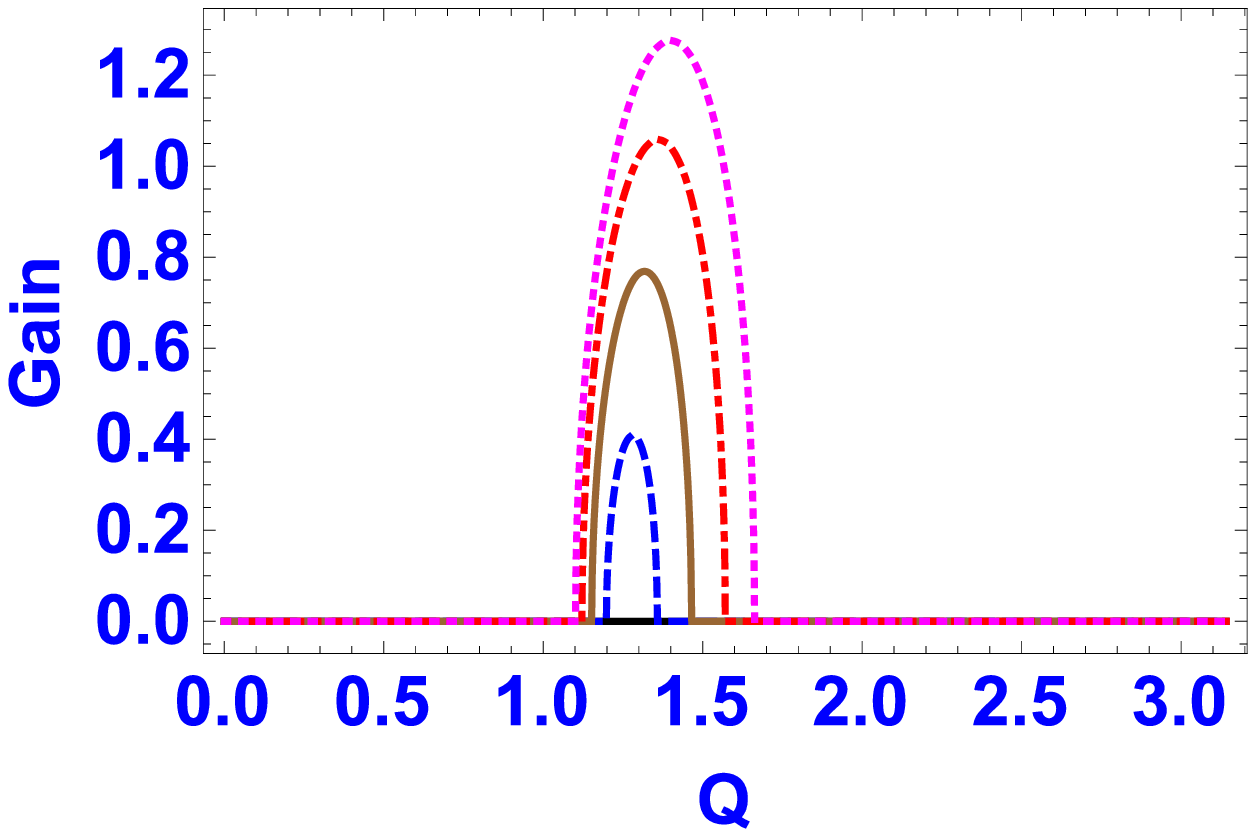}%
\includegraphics[width=.235\textwidth]{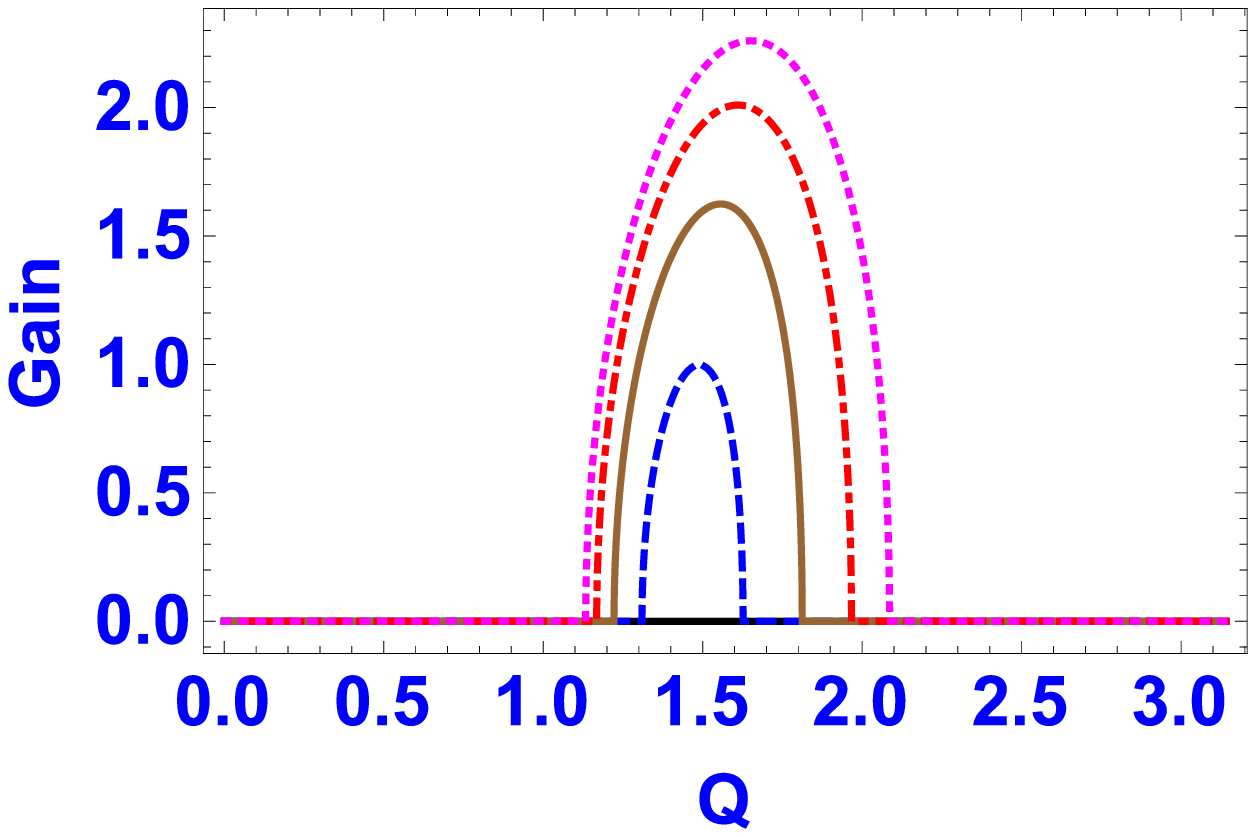}
\caption{(color online) 3D surface plot showing the MI region as a function of $Q$ and $\gamma$ for upper panel: partially miscible, ($g= 1$, $G=1$) and lower row (left): immiscible ($g= 1$, $G=0.5$) and (right): miscible ($g= 1$, $G=1.5$) conditions.}
\label{rf6}
\end{figure}
\begin{figure}[htbp!]
\includegraphics[width=.49\textwidth]{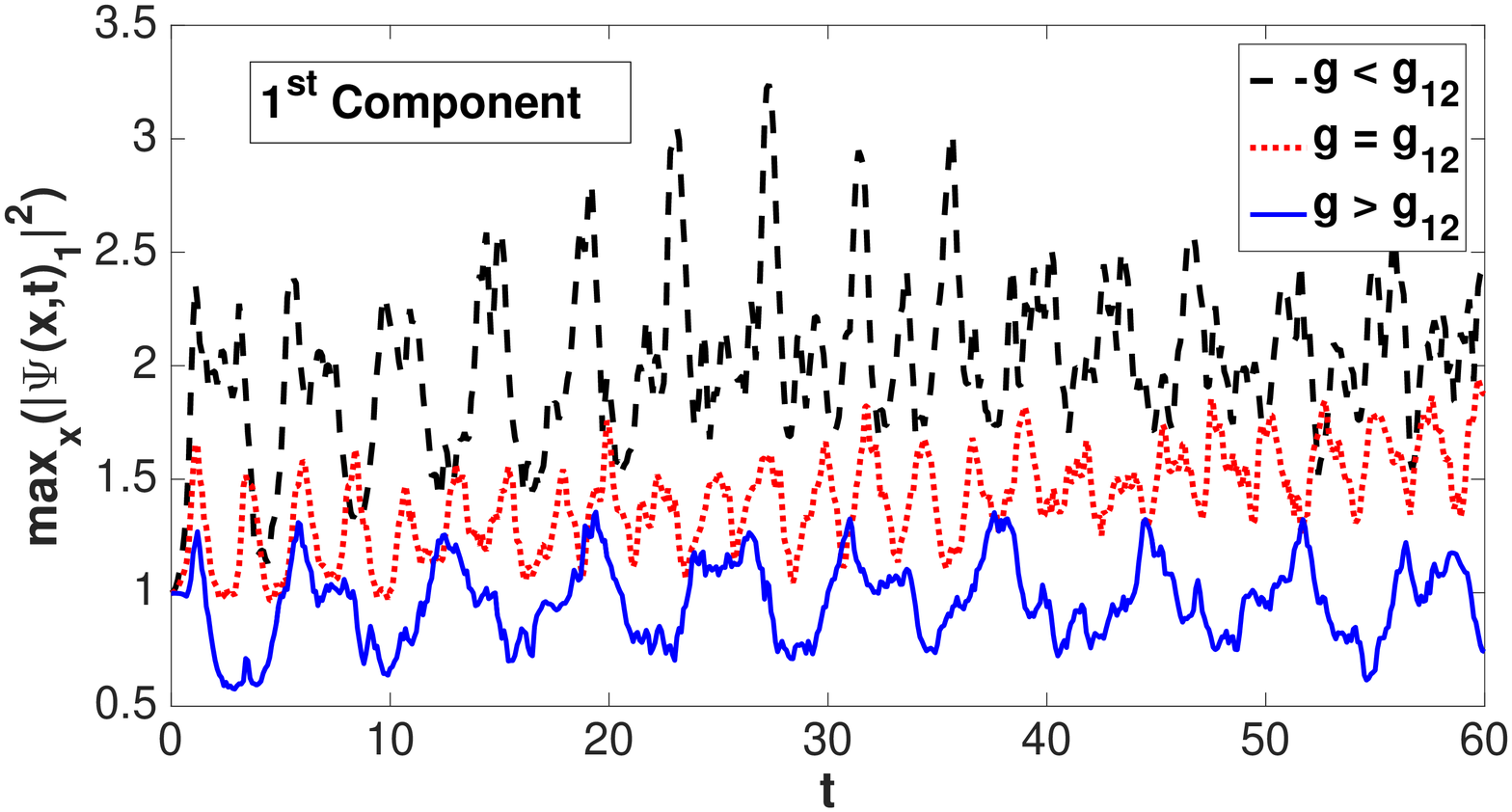}
\includegraphics[width=.49\textwidth]{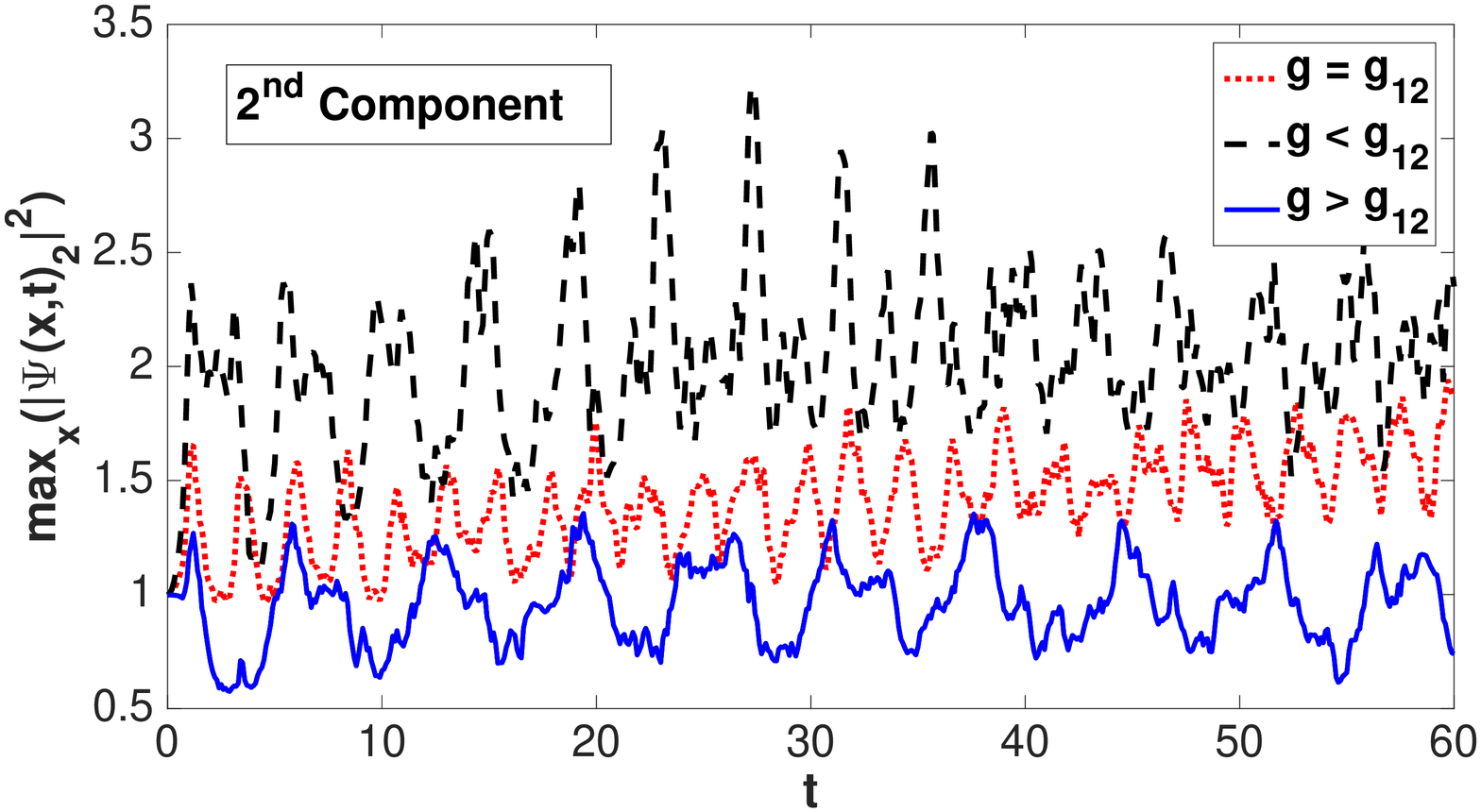}
\caption{Comparative time evolution of the maximum density for miscible, immiscible and semi miscible cases corresponding to the parameters in Fig. 7.}
\label{fnu3}
\end{figure}

Figure~8 shows the comparative evolution of the wave amplitudes in both cases discussed in Fig.~7 for miscible, immiscible and partially miscible cases. Mild and large oscillations of the maximum density illustrate MS and MU cases. 
\color{black}

\begin{figure*}[htbp!]
\includegraphics[width=.235\textwidth]{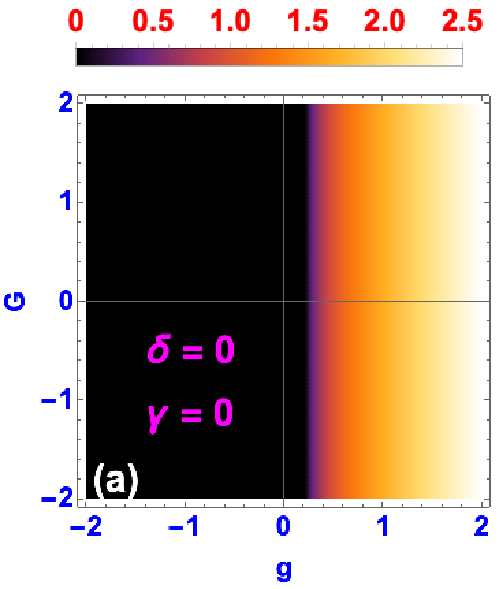}
\includegraphics[width=.235\textwidth]{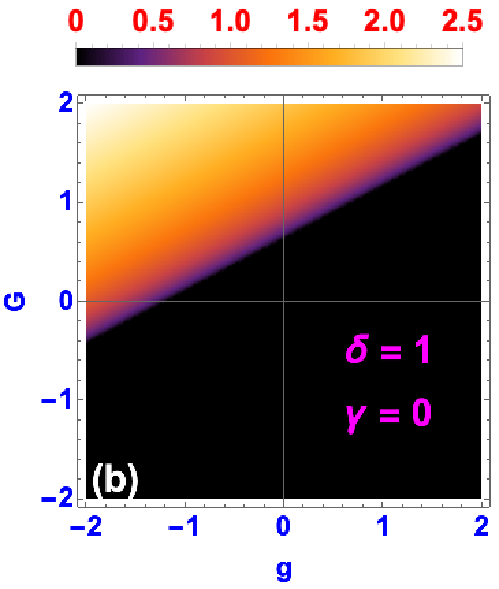}
\includegraphics[width=.235\textwidth]{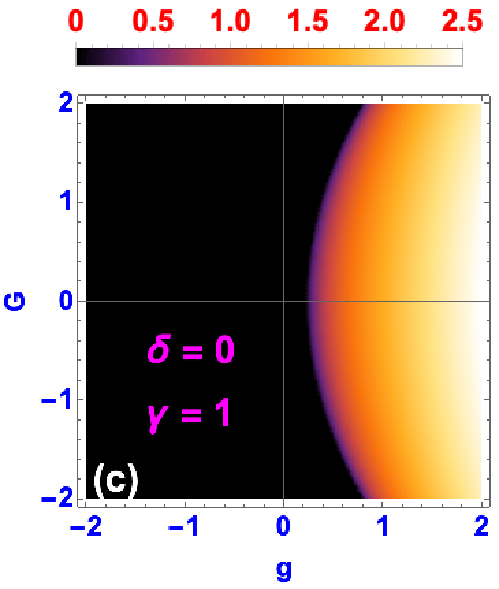}
\includegraphics[width=.235\textwidth]{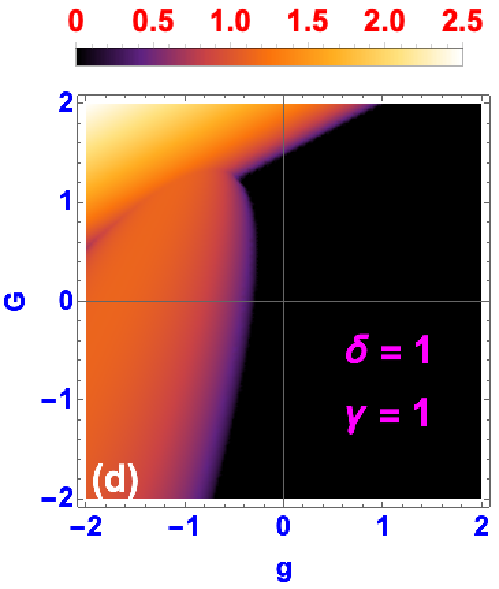}
\caption{(color online) Regions of modulational stability / instability in the $(g,G)$ plane for all possible combinations of Rabi and SOCs.}
\label{f12}
\end{figure*}
\begin{figure*}[htbp!]
\includegraphics[width=.32\textwidth]{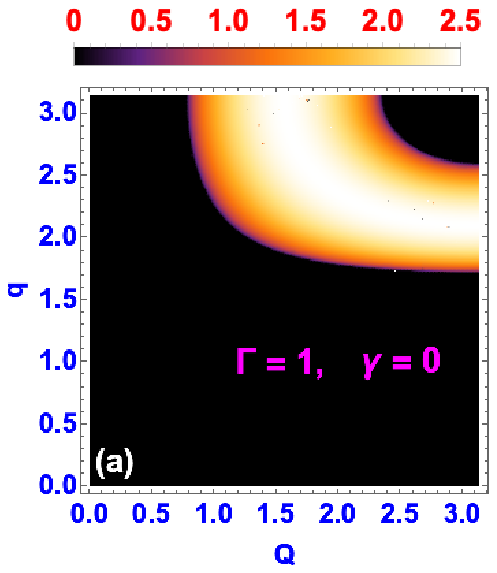}
\includegraphics[width=.32\textwidth]{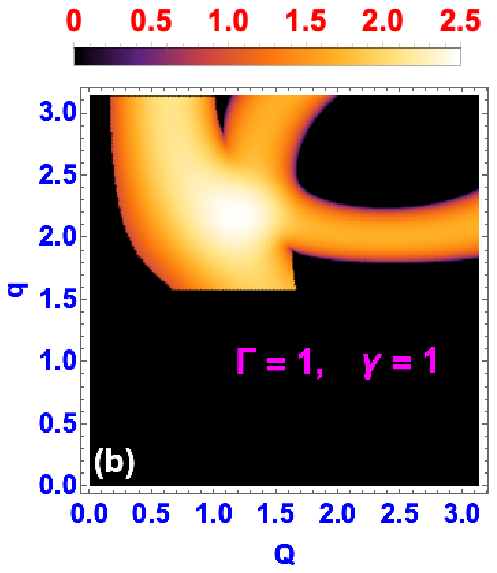}
\includegraphics[width=.32\textwidth]{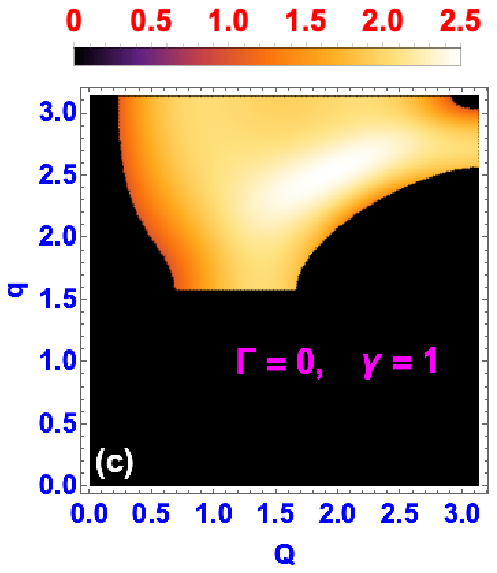}
\caption{(color online) Q Vs q for (a) $\Gamma=1$ and $\gamma=0$, (b) $\Gamma=1$ and $\gamma=1$, (c) $\Gamma=0$ and $\gamma=1$. Remaining parameters are $g=1$, $G=1$ and $\delta=1$.}
\label{f14}
\end{figure*}
In  the absence of both Rabi and SOC, the MI region exists only in the {\it repulsive} two-body inter-atomic interaction zone, as shown in Fig.\ref{f12} (a). 
When we add Rabi coupling alone, the MI region expands from repulsive to attractive two-body inter-atomic as well as intra-atomic interaction zone as shown in Fig.\ref{f12} (b). 
We also observe that in the presence of SOC and absence of Rabi coupling, the MI region exists only in repulsive two-body inter-atomic and both repulsive and attractive intra-atomic interactions zone, as displayed in Fig.\ref{f12} (c). 
Now, when we reinforce SOC along with Rabi coupling, the MI region exists mostly in attractive two-body inter-atomic and both repulsive and attractive intra-atomic interaction zone as shown  in Fig.\ref{f12} (d). 

Further, from panels (a)-(c), where the strength of the Rabi coupling is absent,  the stability zone stays in the same region with small variations. However, from panels (a)-(b), where the strength of the SOC is absent,  we observe that the repulsive inter-atomic region which was stable earlier becomes unstable while repulsive/attractive inter-atomic and intra-atomic regions which were unstable  now shift to stable domain. All the above occurs at the behest of Rabi coupling. 
Hence, from Fig.\ref{f12}, it is pretty obvious that the impact of Rabi coupling is more pronounced compared to the other system parameters and this result underscores the dominance of Rabi coupling. 

In Fig.\ref{f14}, panels (a) and (b) show the effect of SOC in $(q,Q)$ plane. When there is no SOC, the MI regions exist only for $Q\geq 0.8$. 
But, in the presence of SOC, the onset of MI appears at $Q\geq 0.1$. 
In panel (c), we show the effect of the SOC when the hopping strength is absent ($\Gamma=0$). In this case, there is no unstable island inside the stable region as shown in panel (b). In addition, from fig.\ref{f14}, we observe MI region when $q\geq 1.7$.
\section{Conclusion} \label{sec5}
We have investigated the impact of intra-site SO coupling and associated inter-component Rabi coupling on the MI of plane-wave states in two-component discrete BECs. We have brought out the existence of MI domains under the interplay between SOC, Rabi coupling, intra and inter species interactions.  
We have also found that the  SO coupling initiates MI even for a small initial wavenumber in miscible states and also introduces MI even in the absence of hopping coefficient, a concept  which may have wider ramifications in heavy atomic BECs. We have also shown how the results of linear stability analysis can be corroborated numerically.

\section*{Acknowledgements}
\noindent Authors wish to thank the reviewers for their critical comments to improve the contents of the paper. SS wishes to thank the Council of Scientific and Industrial Research(CSIR) for  financial assistance(Grant No. 03(1456)/19/EMR- II), the Government of India. RR wishes to acknowledge the financial assistance from DAE-NBHM(Grant No .02011/3/20/2020/NBHM(R.P)/R $\&$ D II dated 8th September  2020) and CSIR(Grant No 03(1456)/19/EMR-II).


\begin{thebibliography}{99}

\bibitem{Spielman} Y. J. Lin, K. Jimenez-Garcia and I. B. Spielman, Nature \textbf{471} (2011) 83.

\bibitem{Wang} C. Wang, C. Gao, C.M. Jian and H Zhai, Phys. Rev. Lett \textbf{105} (2010) 160403; D.A. Zezyulin, R. Driben, V.V. Konotop and B.A. Malomed, Phys. Rev. A \textbf{88} (2013) 013607.

\bibitem{Li} Y. Li, L.P. Pitaevskii and S. Stringari, Phys. Rev. Lett. \textbf{108} (2012) 225301.

\bibitem{Achilleos} V. Achilleos, D. J. Frantzeskakis, P.G. Kevrekidis and D.E. Pelinovsky, Phys. Rev. Lett. \textbf{110} (2013) 264101.

\bibitem{Kartashov} Y.V. Kartashov, V.V. Konotop and F.K. Abdullaev, Phys. Rev. Lett. \textbf{111} (2013) 060402.

\bibitem{Lobanov} V.E. Lobanov, Y.V. Kartashov and V.V. Konotop, Phys. Rev. Lett. \textbf{112} (2014) 180403.

\bibitem{Xu} Y. Xu, Y. Zhang, B. Wu, Phys. Rev. A. \textbf{87}, 013614 (2013); L. Salasnich and B.A. Malomed, Phys. Rev. A. \textbf{87} (2013) 063625.

\bibitem{Sakaguchi-Li-Malomed1} H. Sakaguchi, B. Li and B.A. Malomed, Phys. Rev. E. \textbf{89} (2014) 032920.

\bibitem{Salasnich-Malomed1} L. Salasnich, W.B. Cardoso and B.A. Malomed, Phys. Rev. A. \textbf{90} (2014) 033629.

\bibitem{Sakaguchi-Li-Malomed2} H. Sakaguchi and S. Maeyama, Phys. Rev. E. \textbf{87} (2013) 024901.
\bibitem{Sakaguchi-Meqs} H. Sakaguchi and B.A. Malomed, Phys. Rev. E. \textbf{90} (2014) 062922.

\bibitem{YZhang} Y. Zhang and C. Zhang, Phys. Rev. A. \textbf{87} (2013) 023611.

\bibitem{Larson} J. Larson, J. P. Martikainen, A. Collin and E. Sjoqvist, Phys. Rev. A. \textbf{82} (2010) 043620.

\bibitem{Stanescu} T. D. Stanescu, V. Galitski, J. Y. Vaishnav, C. W. Clark and S. Das Sarma, Phys. Rev. A. \textbf{79} (2009) 053639.


\bibitem{Trombettoni} A. Trombettoni and A. Smerzi, Phys. Rev. Lett. \textbf{86} (2001) 2353.

\bibitem{Kevrekidis} P. G. Kevrekidis, The Discrete Nonlinear Shoringer Equation: Mathematical Analysis, Numerical Computations and Physical Perspective(Berlin:Springer) 2009.

\bibitem{Gligoric} G Gligoric, A. Maluckov, L.Hadzievski and B. A. Malomed, Phys. Rev. A. \textbf{78} (2008) 063615; Phys. Rev. B \textbf{88} (2013) 155329; Chaos \textbf{24} (2014) 023124.

\bibitem{Belicev} P.P. Belicev, G. Gligoric, J. Petrovic, A. Maluckov, L.Hadzievski and B.A. Malomed, J. Phys. B: At. Mol. Opt. Phys. \textbf{11} (2015) 065301.

\bibitem{Benjamin} T. B. Benjamin, J. E. Feir, J. Fluid. Mech. \textbf{27} (1967) 417; L. A. Ostrovskii, Sov. Phys. Jetp. \textbf{24} (1969) 797; A. Hesegawa, Phys. Rev. Lett. \textbf{24} (1970) 1165.

\bibitem{Al Khawaja} U. Al Khawaja, H. T. C. Stoof, R. G. Hulet, K.E. Strecker, and G. B. Partridge, Phys. Rev. Lett. \textbf{89} (2002) 200404.

\bibitem{K.E.Strecker} K.E. Strecker, G. B. Partridge, A. G. Truscott, and R. G. Hulet, Nature (London) \textbf{417}, 150 (2002); L. D. Carr and J. Brand, Phys. Rev. Lett. \textbf{92} (2004) 040401.

\bibitem{Sabari} S. Sabari, E. Wamba, K. Porsezian, A. Mohamadou, and T. C. Kofane, Phys. Lett. A \textbf{377} (2013) 2408; E. Wamba, S. Sabari, K. Porsezian, A. Mohamadou, and T. C. Kofane, Phys. Rev. E \textbf{89} (2014) 052917; S. Sabari, K. Porsezian, and R. Murali, Phys. Lett. A \textbf{379} (2015) 299.

\bibitem{Goldstein} E. V. Goldstein and P. Meystre, Phys. Rev. A \textbf{55} (1997) 2935.
\bibitem{Kasamatsu1} K. Kasamatsu and M. Tsubota, Phys. Rev. Lett. \textbf{93} (2004) 100402.
\bibitem{Kasamatsu2} K. Kasamatsu and M. Tsubota, Phys. Rev. A \textbf{74} (2006) 013617.

\bibitem{Kobyakov 2C MI} A. Kobyakov, S. Darmanyan, F. Lederer, and E. Schmidt, Opt.Quantum Electron. \textbf{30} (1998) 795.
\bibitem{ZRaptiMI} Z. Rapti, A. Trombettoni, P .G. Kevrekidis, D. J. Frantzeskakis, B. A. Malomed, A. R. Bishop, Phys. Lett. A. \textbf{330} (2004) 95.

\bibitem{Baizakov 2C MI} B. B. Baizakov, A. Bouketir, A. Messikh and B.A.Umarov, Phys. Rev. E. \textbf{79} (2009) 046605.

\bibitem{Guang 2C MI} Guang-Ri Jin, Chul Koo Kim and, Kyun Nahm, Phys. Rev. A \textbf{72} (2005) 045601.

\bibitem{Ruostekoski} J. Ruostekoski and Z. Dutton, Phys. Rev. A \textbf{76} (2007) 063607.

\bibitem{ishfaq}I.A. Bhat, T. Mithun, B.A. Malomed and K. Porsezian, Phys. Rev. A \textbf{92} (2015) 063606.

\bibitem{mithun} T. Mithun1 and K. Kasamatsu, J. Phys. B: At. Mol. Opt. Phys. \textbf{52} (2019) 045301.

\end{thebibliography}
\end{document}